\documentclass[a4paper,fleqn,usenatbib]{mnras}

\usepackage{newtxtext,newtxmath}

\usepackage[T1]{fontenc}
\usepackage{hyperref}
\DeclareRobustCommand{\VAN}[3]{#2}
\let\VANthebibliography\thebibliography
\def\thebibliography{\DeclareRobustCommand{\VAN}[3]{##3}\VANthebibliography}

\usepackage{graphicx}	
\usepackage{amsmath}	

\title[Extraplanar gas in UGCA\,250]{Uncovering Extraplanar Gas in UGCA\,250 with the Ultra-deep MHONGOOSE Survey}

\author[Kurapati et al.]{
Sushma Kurapati$^{1,2}$\thanks{E-mail: kurapati@astron.nl (SK)},
D. J. Pisano$^{1}$,
 W. J. G. de Blok$^{2,1,3}$,
Peter Kamphuis$^{4}$,
Nikki Zabel$^{1}$, 
Mikhail de Villiers$^{1,5}$,
\newauthor
Julia Healy$^{2, 6, 7}$,
Filippo M. Maccagni$^{8}$,
Dane Kleiner$^{2}$,
Elizabeth A. K. Adams$^{2,3}$,
Philippe Amram$^{9}$,
\newauthor
E. Athanassoula$^{9}$, 
Frank Bigiel$^{10}$,  
Albert Bosma$^{9}$,
Elias Brinks$^{11}$,
Laurent Chemin$^{12, 13}$,
Francoise Combes$^{14}$,
\newauthor
Ralf-Jürgen Dettmar$^{4}$,
Gyula Józsa$^{15, 16}$, 
Baerbel Koribalski$^{17, 18}$,
Antonino Marasco$^{19}$,
Gerhardt Meurer$^{20}$, 
\newauthor
Moses Mogotsi$^{5, 21, 1}$,
Abhisek Mohapatra$^{1}$,
Sambatriniaina H. A. Rajohnson$^{1}$,
Eva Schinnerer$^{22}$, 
\newauthor
Amidou Sorgho$^{23}$, 
Kristine Spekkens$^{24, 25}$, 
Lourdes Verdes-Montenegro$^{23}$,
Simone Veronese$^{2,3}$, 
 Fabian Walter$^{22}$\\
(Affiliations can be found after the references)
}
\date{Accepted XXX. Received YYY; in original form ZZZ}
\pubyear{2024}
\begin{document}
\label{firstpage}
\pagerange{\pageref{firstpage}--\pageref{lastpage}}
\maketitle

\newcommand{\HI}{\rm H{\sc i }}
\newcommand{\MB}{\ensuremath{\rm M_B}}
\newcommand{\mb}{\ensuremath{\rm M_{bar}}}
\newcommand{\mhi}{\ensuremath{\rm M_{HI}}}
\newcommand{\mg}{\ensuremath{\rm M_{g}}}
\newcommand{\ms}{\ensuremath{\rm M_{s}}}
\newcommand{\msun}{\ensuremath{\rm M_{\odot}}}
\newcommand{\fatm}{\ensuremath{\rm f_{atm}}}
\newcommand{\kms}{\ensuremath{\rm km \, s^{-1}}}
\newcommand{\vhel}{\ensuremath{ V_{\rm hel}}}
\newcommand{\kpc}{\ensuremath{\rm kpc}}
\newcommand{\Jykms}{\ensuremath{ \rm Jy \, km \, s^{-1} }}
\newcommand{\mJybeam}{\ensuremath{ \rm mJy \, beam^{-1} }}
\newcommand{\LCDM}{\ensuremath{\Lambda{\rm CDM}}}
\newcommand{\fat}{{\sc fat }}
\begin{abstract}
We use the neutral atomic hydrogen (H{\sc i}) observations of the edge-on galaxy UGCA\,250, taken as part of the MeerKAT \HI Observations of Nearby Galactic Objects - Observing Southern Emitters (MHONGOOSE) survey to investigate the amount, morphology, and kinematics of extraplanar gas. The combination of high column density sensitivity and high spatial resolution of the survey over a large field of view is ideal for studying the underlying physics governing the extraplanar gas. These data reveal 9 additional detections within the field of view along with UGCA\,250, with 8 of them being within $\sim$ 200 km s$^{-1}$ of the galaxy’s systemic velocity. The galaxy seems to have a tail-like feature extending away from it in the southern direction  up to $\sim$ 41 kpc (in projection). We also detect a cloud at anomalous velocities, but we did not find any optical counterpart. We construct a detailed tilted ring model for this edge-on galaxy to gain a deeper understanding of the vertical structure of its neutral hydrogen. The model that best matches the data features a thick disc with a scale height of $\sim$ 3$\pm$1 kpc and an \HI mass of about 15$\%$ of the total H{\sc i} mass. This extraplanar gas is detected for the first time in UGCA\,250. Our analysis favours a mixed origin for the extraplanar gas in UGCA 250, likely arising from a combination of internal stellar feedback and external tidal interactions.
\end{abstract}



\section{Introduction}

It is becoming increasingly clear that gas accretion is necessary to sustain the observed levels of star formation rates in disk galaxies. Gas accretion onto galaxies can occur via different mechanisms, such as wet mergers with gas-rich satellites \citep{Rubin12} or gradual cooling of gas accreted onto the galactic halo from the cosmic web \citep{vandevoort11, wetzel15, Cornuault18, Ho19, Iza22}, possibly triggered by hydrodynamical interactions with the galactic fountain \citep[e.g.][]{Fraternali02, Melioli08, Melioli09, Marinacci10, Fraternali17}. The halo of a galaxy will form its connection with the intergalactic medium (IGM) and one must focus on the disc-halo interface region to understand the gas exchange processes between the disc and the circum galactic medium (CGM).

 Deep H{\sc i} observations of several nearby spiral galaxies have shown the presence of cold extraplanar gas (EPG) in a gaseous layer that surrounds the star forming disc up to a height of some kpc \citep[e.g.][]{Sancisi08, Hess09, Lucero15}. This layer typically has a mass of 10 -- 30 percent of the mass of the H{\sc i} in the disc \citep[e.g.][]{Oosterloo07, Marasco19}.  
 Observations of nearby spiral galaxies with different inclination angles provide information on the structure and kinematics of neutral EPG in disc galaxies (see \citet{Boomsma08} for face-on galaxies, \citet{Fraternali01, Hess09, Marasco19} for intermediate inclination galaxies, and \citet{Oosterloo07, Rand08, Zschaechner11} for edge-on galaxies).

For example, the Westerbork Synthesis Radio Telescope (WSRT) observations  of the edge-on galaxy NGC\,891 revealed a slowly rotating (also known as lagging) H{\sc i} halo, which extends
vertically out to 14 kpc from the thin \HI disc and, in one quadrant up to 22 kpc, with filamentary structures far from the disc \citep{Swaters97, Oosterloo07}. In the face-on galaxy NGC\,6946, vertical motions are observed across the star forming disc and are in clear connection with EPG \citep{Boomsma08}. In some galaxies, non-circular motions have also been detected. The extraplanar gas of NGC\,2403, UGCA\,105, NGC\,4559 have a coherent radial infall motion towards the centre of the galaxy \citep{Fraternali02, Barbieri05, Schmidt14, Vargas17}. Trends of EPG, such as the presence and magnitude of any decrease in rotation speed with height are useful in tracing their origin. 

Various mechanisms have been proposed to explain the origin of EPG. One is an internal mechanism in which a large part of this halo gas is brought up from the disc by galactic fountains \citep[e.g.][]{Shapiro76, Bregman80}. In this model, supernova explosions expel the hot gas to large distances from the disc into the halo. In the halo the gas cools down and condenses before returning to the disc. The other is an external mechanism in which a galaxy accretes gas from the CGM \citep{Binney05, Kaufmann06}. Additionally, a ``mixed'' scenario has been proposed where the EPG is formed by the interplay between the galactic fountain and the hot CGM. The interactions between these two components produce a mixture with low cooling timescales and slow rotational speed, which explains the observed EPG kinematics and provides a potential source of pristine gas accretion onto the galaxy \citep{Fraternali08, Marasco12, Li23}. A powerful method to determine the origin of EPG is to study their kinematics. In particular, edge on galaxies provide a great opportunity to measure the vertical distribution of the interstellar medium. 

Here, we use the ``full depth'' H{\sc i} 21 cm observations of the nearly edge-on galaxy UGCA\,250 (J1153-28), taken as part of the MeerKAT \HI Observations of Nearby Galactic Objects - Observing Southern Emitters (MHONGOOSE) survey \citep{deBlok24} to  study the detailed spatial and kinematical distribution of the gas, with a focus on the faint outer edges. UGCA\,250 is a nearby galaxy at a distance of 20.4 Mpc \citep{Kourkchi22}. The galaxy has an optical diameter (d$_{25}$) of 21.6 kpc, stellar mass log(M$_{\ast}$/ M$_{\odot}$) = 9.80, and star formation rate of 0.54 M$_{\odot}$ yr$^{-1}$, which places it on the star formation main sequence for spiral galaxies.  Table \ref{tab:HI_properties} lists the basic parameters of UGCA\,250. The MHONGOOSE survey is an ultra-deep H{\sc i} survey that uses the MeerKAT radio interferometer to observe the H{\sc i} in 30 nearby galaxies. The combination of high column density sensitivity and high spatial resolution over a large field of view is ideal for studying the underlying physics governing the EPG as well as the disc–halo connection. 

 This paper is organized as follows. In \S \ref{sec:obs}, we describe the observations and data reduction techniques. In \S \ref{sec:results}, we presents the results, including the detection of companion galaxies, \HI morphology of UGCA\,250, and tilted ring models. In \S \ref{sec:disc}, we discuss the various models used to understand the origin of the extraplanar gas. Finally, in \S \ref{sec:summary}, we summarize our key findings.

\begin{table}
  \begin{center}
    \caption{General properties of UGCA\,250.}
    \begin{tabular}{cc}
      \hline
      \hline
      $\alpha, \delta$ (J2000) &  $\rm 11^{h}53^{m}24.1^{s}$, $-28^{\circ}33^{'}11.4^{''}$ \\
      V$_{\rm sys}$ (\kms) & 1705$^{a}$ \\
      PA ($^{\circ}$) & 256$^{a}$\\
      $i$ ($^{\circ}$) & 82$^{a}$ \\
      $d$ (Mpc) & 20.4$^{b}$ \\
      log(M$_{\ast}$/ M$_{\odot}$) & 9.80$^{c}$ \\
      SFR (M$_{\odot}$/ year) & 0.54$^{c}$ \\
      D$_{\rm 25}$ (kpc) & 21.6$^{c}$ \\
      D$_{\rm HI}$ (kpc) & 69.0$^{a}$ \\
      \hline
    \end{tabular} \\
    \smallskip
    \footnotesize Notes: $^{a}$From this work (at \HI column denisty $\sim$ 1.2$\times$10$^{19}$ cm$^{-2}$); $^{b}$Distance is taken from \citet{Kourkchi22}; $^{c}$From \citet{Leroy19}; $\alpha, \delta$ (J2000): Coordinates; V$_{\rm sys}$: Heliocentric systemic velocity; PA: Position angle; $i$: Inclination angle; $d$: Distance; log(M$_{\ast}$/ M$_{\odot}$): Stellar mass; SFR: Star formation rate; D$_{\rm 25}$: Optical diameter; D$_{\rm HI}$: \HI diameter.
    \label{tab:HI_properties}
  \end{center}
\end{table}




\begin{figure}
    \centering
    \includegraphics[width=0.98\linewidth]{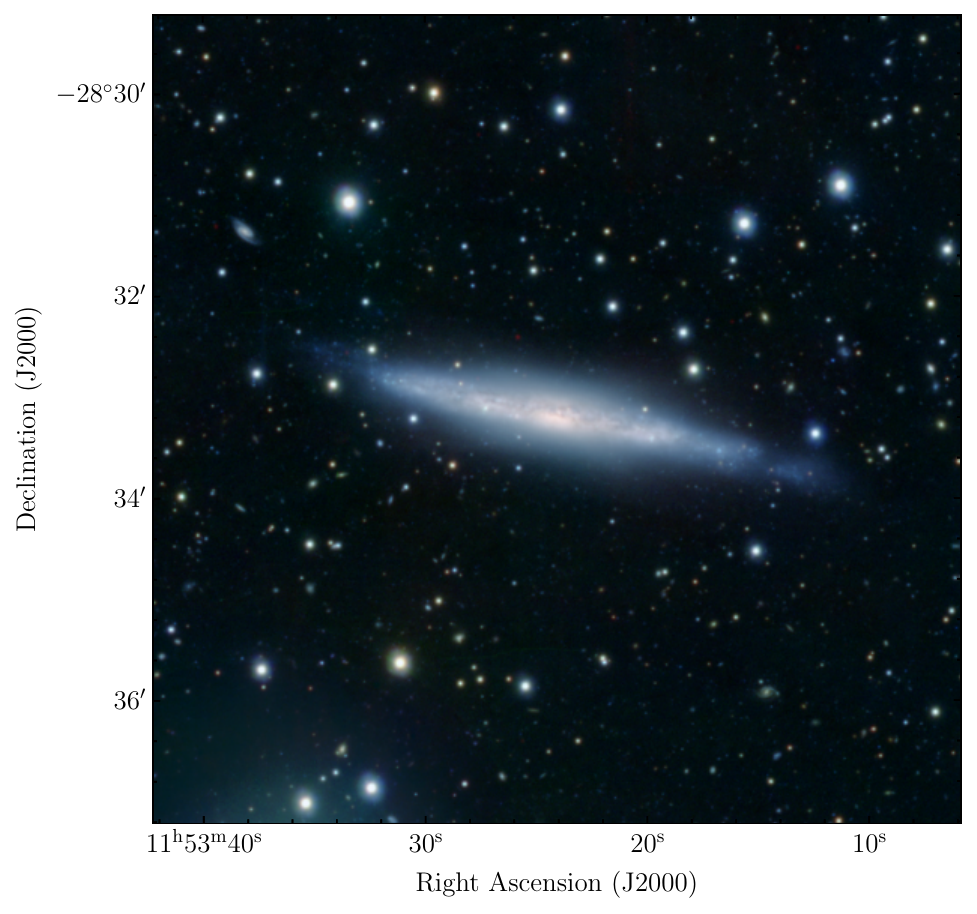}
    \caption{Composite optical image UGCA\,250. The RGB colour is provided by a combination of $g$, $r$, and $i$-band VST data. At brighter magnitudes, the galaxy exhibits a thin, elongated structure typical of an edge-on galaxy. At fainter magnitudes, it shows a vertical extension and appears asymmetric, with an extension towards the southwest.}
    \label{fig:optical}
\end{figure}


\begin{figure*}
    \centering
    \includegraphics[width=0.99\linewidth]{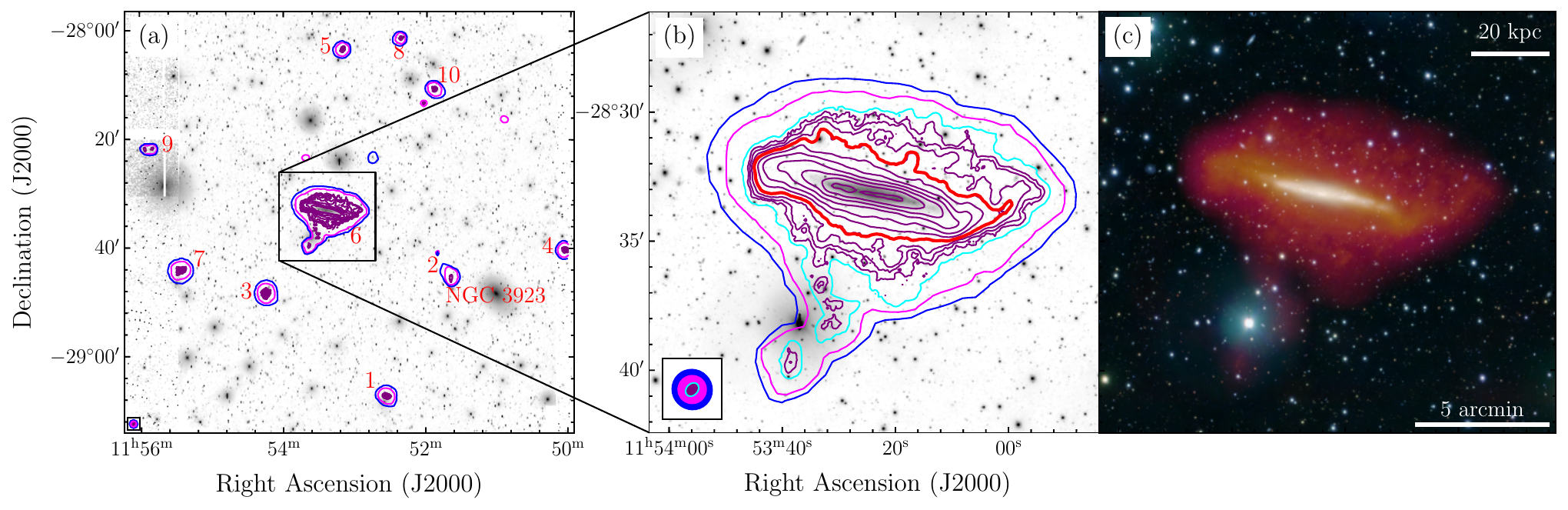}
    \caption{. The \HI integrated intensity contours are overlaid on the VST optical $g$-band image. (a) Moment 0 map of UGCA\,250 and its 9 satellite galaxies. The contours are shown from data at various resolutions: 26.7\arcsec $\times$ 18.8\arcsec ({\tt r10t00}) in purple with levels of 1.2$\times$10$^{19}$ cm$^{-2} \times 2^{n}$, n = 0,1,2,3 ..; 34.3\arcsec $\times$ 25.6\arcsec ({\tt r15t00}) in cyan at 3.0$\times$10$^{18}$ cm$^{-2}$; 65.3\arcsec $\times$ 63.8\arcsec ({\tt r05t60}) in magenta at 1.0$\times$10$^{18}$ cm$^{-2}$; and 93.8\arcsec $\times$ 91.7\arcsec ({\tt r10t90}) in blue at 6.2$\times$10$^{17}$ cm$^{-2}$. (b) Moment-0 map of the galaxy UGCA 250 with the same \HI contours, where the thick red contour corresponds to 1.0 $\times$10$^{20}$ cm$^{-2}$. Note that the \HI\ column density values have not been corrected for the primary beam attenuation.  The ellipses in the bottom left corner represent the beams at each resolution. The zoomed-in moment 0 and moment 1 maps of individual satellite galaxies are shown in Figs \ref{fig:sat1} and \ref{fig:sat2} in the appendix. (c) The red, orange, and yellow show the \HI integrated intensity from three different resolution data cubes: 14.4\arcsec $\times$ 9.8\arcsec, 26.7\arcsec $\times$ 18.8\arcsec, and 65.3\arcsec $\times$ 63.8\arcsec. These H{\sc i} maps are overlaid on a composite optical RGB image of UGCA\, 250 in the background. The creation of this image followed the technique outlined in \citet{English2017}.} 
    \label{fig:mom0_overlay}
\end{figure*}

\begin{table*}
\caption{HI properties of satellite galaxies }
\begin{tabular}{cccccrrcc}
\hline
ID & Galaxy ID & RA & Dec & $V_{\rm sys}$ & $W_{\rm 20}$  & $W_{\rm 50}$ & Flux & \HI mass  \\
    &  & (h m s) & ($^{\circ}$ $^{\prime}$ $^{\prime\prime}$)& (km s$^{-1}$) & (km s$^{-1}$) & (km s$^{-1}$) &  (Jy km s$^{-1}$) &  log($M_{\rm HI}$/$M_{\odot}$)  \\
\hline
 1 &ESO 440- G 023$^{\ast}$ & 11 52 33.1 & -29 07 19.1 & 1866 & 114.4 & 92.3 &  0.31 & 7.49 \\
2 & 2MFGC 09308 & 11 51 39.3 & -28 45 17.5  & 1887 & 71.6 & 46.8 &  0.04 & 6.60 \\
3 & LEDA 737259$^{\ast}$ & 11 54 14.8 & -28 48 19.8 & 1761 & 42.7 & 27.5 & 0.62 & 7.79 \\
4 & ESO 440- G 014$^{\ast}$  & 11 50 03.9 & -28 40 18.2 & 1857 & 68.9 & 44.1 &  0.10 & 6.98 \\
5 & LEDA 746393$^{\ast}$ & 11 53 10.4 & -28 03 30.4 & 1828 & 95.1 & 71.6 & 0.09 & 6.96 \\
6 & UGCA\,250$^{\ast}$ & 11 53 24.3 & -28 33 08.9 & 1694 & 268.7 & 285.3 & 72.0 & 9.85 \\
7 & ESO 440- G 030$^{\ast}$ & 11 55 25.9  & -28 44 10.2 & 1801 & 115.8 & 91.0  & 0.78 & 7.89 \\
8 & MKT115222-280126 & 11 52 22.1 & -28 01 26.2 & 1636 & 42.7 & 28.9 & 0.02 & 6.30 \\
9 & MKT115551-282149 & 11 55 51.4  & -28 21 49.3 & 1595 & 68.9 & 45.5 & 0.01 & 6.10 \\
10 & 6dFJ1151533-281047$^{\ast}$ & 11 51 53.3 & -28 10 50.2 & 1422 & 78.5 & 50.9 & 0.08 & 6.89 \\
\hline
 & Cloud at anomalous velocities (C) & 11 53 20.5 & -28 35 47.0 & 1605 & -- & -- & 0.065 & 6.81 \\ 
 & Blob below filament (B) & 11 53 38.4 & -28 39 43.0 & 1640 & -- & -- & 0.02 & 6.34 \\
   & Entire Filament (with blob) & 11 53 31.3 & -28 37 34.0 & 1617 & -- & -- & 0.10 & 7.01 \\
\hline
\multicolumn{9}{l}{$^{\ast}$ Galaxies that were previously detected in ``single track'' (5.5hr) shallow data \citep{deBlok24}; ID: Identifier; Optical ID: Optical identification;} \\
\multicolumn{9}{l}{ RA: Right Ascension; Dec: Declination; $V_{\rm sys}$: Systemic velocity in km s$^{-1}$; $W_{\rm 20}$ and $W_{\rm 50}$: \HI line width at 20\% peak and line width at 50\% peak in km s$^{-1}$;} \\
\multicolumn{9}{l}{ Flux: Integrated flux in Jy km s$^{-1}$; \HI mass: \HI mass in solar masses ($M_{\odot}$).} \\
\label{table:satellites}
\end{tabular}
\end{table*}

\section{Observations and Data reduction}
\label{sec:obs}

The MeerKAT observations of UGCA 250 were taken as part of the MHONGOOSE survey in 10 tracks each lasting 5.5 hours, resulting in a total on source time of 50 hours. Each track consists of 10 min of observing time on one of the primary calibrators, J0408-6545 or J1939-6342, and the secondary calibrator, J1154-3505, was observed every 50 min for 2 min. We used the narrow-band mode of MeerKAT which has 32768 channels of 3.265 kHz each, giving a total bandwidth of 107 MHz. This corresponds to a rest frame velocity resolution of 0.7 \kms at 1420 MHz. All data were subsequently binned to 1.4 \kms for the reduction process.

We use the Containerised Automated Radio Astronomy Calibration (\texttt{CARACal}\footnote{\href{https://github.com/caracal-pipeline/caracal}{https://github.com/caracal-pipeline/caracal}}) pipeline \citep{Jozsa20} to carry out the standard calibration and data reduction processes.  \texttt{CARACal} is based on \texttt{STIMELA}, a radio interferometry scripting framework based on container technologies and \texttt{Python} \citep{Makhathini18}, which uses many publicly available radio interferometry software packages such as 
\texttt{CASA}, \texttt{Cubical}, and \texttt{WSClean}, etc. The pipeline flags the calibrators for radio frequency interference, derives and applies the cross-calibration solutions, and splits off the target. Further, we use the self-calibration of the continuum to improve the quality of the calibration. After applying the self-calibration, we subtract the continuum from the measurement set. 

The \HI cubes were then imaged at various resolutions using \texttt{WSClean} as part of the \texttt{CARACal}. We make six \HI cubes that range in resolution from 6$''$ to 90$''$ using different combinations of robust weighting (r) and Gaussian tapering (t), leading to varying \HI column density sensitivity ranging from N$_{\rm HI} \sim$ $6.0 \times 10^{19} $ cm$^{-2}$ to $5.7 \times 10^{17} $ cm$^{-2}$ at 3$\sigma$ over 16 \kms. For most of the analysis, we use {\tt r15t00} configuration, where $r = 1.5$ and $t = 0$. This yields a resolution of 34.3$\arcsec \times$ 25.6$\arcsec$ ($\sim$ 3 kpc at the distance of UGCA\,250), and achieves a 3$\sigma$ column density sensitivity of approximately $2.7 \times 10^{18}$ cm$^{-2}$ over 16 \kms, which offers an optimal balance between column density sensitivity and spatial resolution. A more detailed description of the data reduction and data quality assessment is presented in \citet{deBlok24}.

We use {\tt SoFiA-2} \citep{Westmeier21} to search for sources and to generate the moment maps of the detections, which are standard maps produced for all the MHONGOOSE observations. We smooth the cube spatially and spectrally with the `smooth and clip' algorithm of {\tt SoFiA-2}. We apply a threshold of 4 $\sigma$ using two spatial Gaussian smoothing kernels: no smoothing, one times the beam;  in addition to three spectral boxcar smoothing kernels: no smoothing, 12 \kms, and 35 \kms.  We then create moment maps of detections from the masked data cubes and we correct the moment-0 maps for the primary beam attenuation of MeerKAT \citep[see][for more details]{deBlok24, Healy24}.

\begin{figure*}
    \centering
    \includegraphics[width=0.99\linewidth]{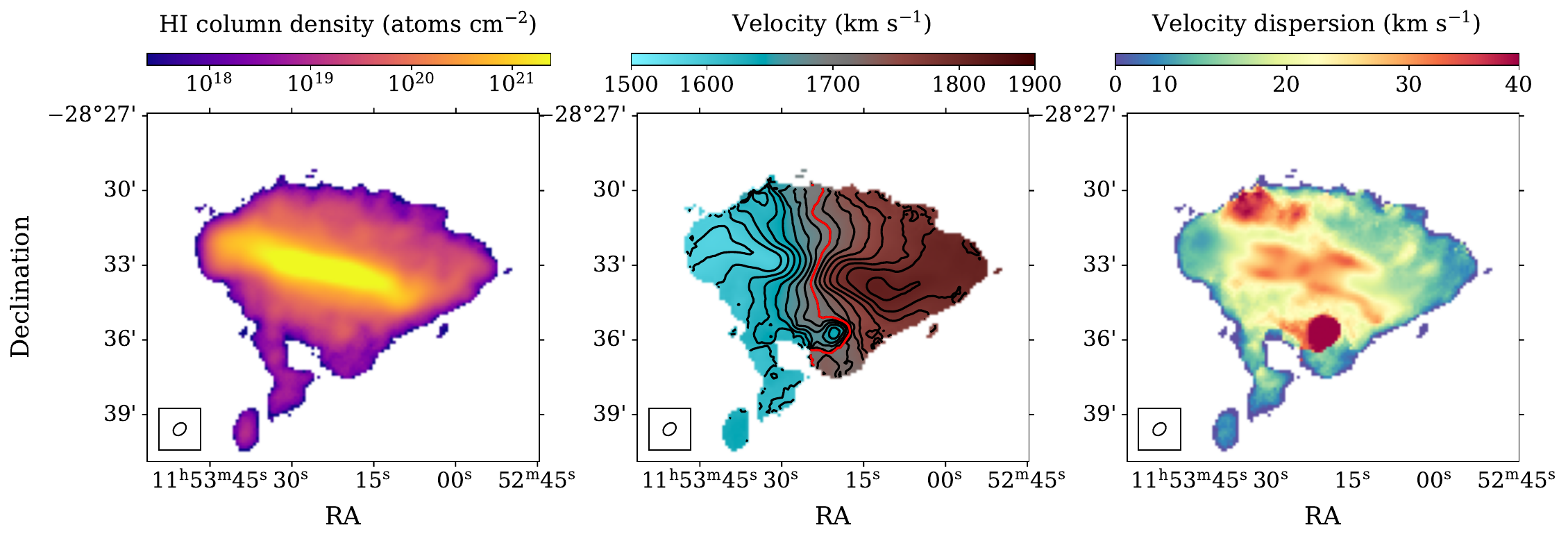}
    \caption{Moment maps of UGCA\,250 at a resolution of 34.3$\arcsec$ $\times$ 25.6$\arcsec$. Left panel: The \HI intensity map (moment 0). Middle panel: velocity field (moment 1),  The isovelocity contours are spaced at 15 \kms and the systemic velocity is indicated by a red line. The color scheme used in this panel is developed by \citet{English24}. Right panel: observed line width/ dispersion map of the galaxy, UGCA\,250.}
    \label{fig:UGCA250_moment_maps}
\end{figure*}

In this paper, we use deep optical observations made using OMEGACAM on the VLT Survey Telescope (VST) for UGCA\,250 from dedicated proprietary observations (ID: 110.25AJ, PI: Maccagni; ID: 112.266Y, PI: Marasco) of the MHONGOOSE galaxies (hereafter MHONGOOSE--VST).
These observations reach a limiting surface brightness of $\mu_{g} = 28.5$ mag/arcsec$^{2}$, $\mu_{r} = 27.3$ mag/arcsec$^{2}$ in the g and r bands respectively, for a point source at 5 $\sigma$ over a circular area with FWHM $\sim$ 1$^{''}$. To create the RGB image shown in Fig. \ref{fig:optical}, we combine the $g$- and $r$-bands from MHONGOOSE-VST with publicly available $i$-band observations from the VST Early-type Galaxy Survey (VEGAS; \citealt{Spavone17, Iodice21}). Although VEGAS primarily targets elliptical galaxies, UGCA\,250 was observed because it lies approximately 30$\arcmin$ away from the elliptical galaxy NGC\,3923 \citep{Kourkchi17}.
At brighter magnitudes, the optical morphology resembles an edge-on galaxy, characterized by a thin, elongated structure. However, at fainter magnitudes, it exhibits a vertical extension, and the galaxy appears asymmetric with a with a noticeable discontinuity towards the southwestern side. This discontinuity is more likely a feature connected to the main disk rather than a separate dwarf galaxy. As will be discussed in Section \ref{sec:HI_morphology}, H{\sc i} observations do not reveal emission at a  distinct velocity at that location, a signature expected if a separate dwarf were present.
Since the MHONGOOSE-VST observations do not fully cover the surrounding dwarf galaxies, we instead use the VEGAS $g$-band image in Fig. \ref{fig:mom0_overlay}, which provides a more complete view of the environment around UGCA\,250, including all nearby dwarfs.


\begin{figure}
    \centering
    \includegraphics[width=0.90\linewidth]{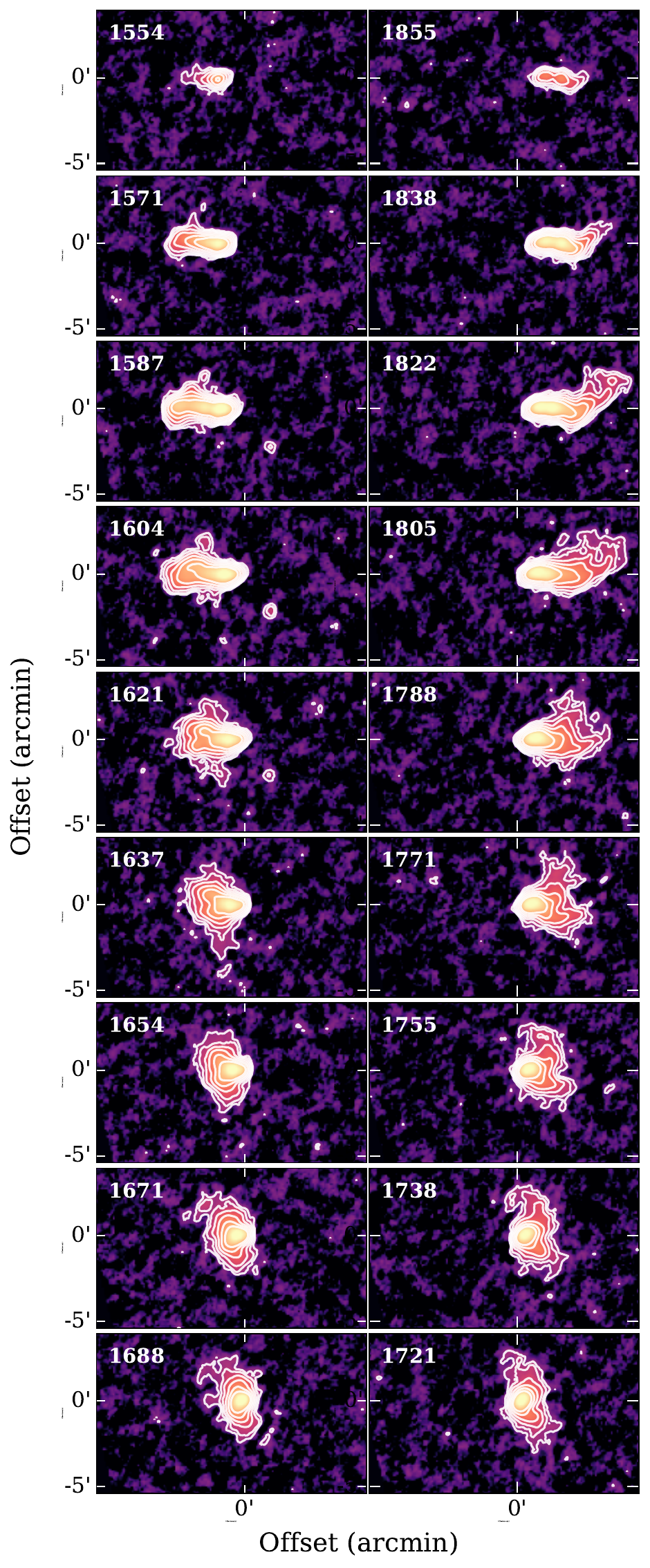}
    \caption{Selected channel maps of UGCA\,250 at the resolution of 34.3$\arcsec$ $\times$ 25.6$\arcsec$ ({\tt r15t00}, $r =$ 1.5 and $t =$ 0). The heliocentric velocity in \kms is indicated on top-left of each panel. Contours begin at 3$\sigma$ and increase by factors of 2. The presence of cloud and filament is evident in the channels at velocities ranging from 1587 to 1637 km s$^{-1}$} 
    \label{fig:channel_maps}
\end{figure}

 \begin{figure}
    \centering
    \includegraphics[width=1.0\linewidth]{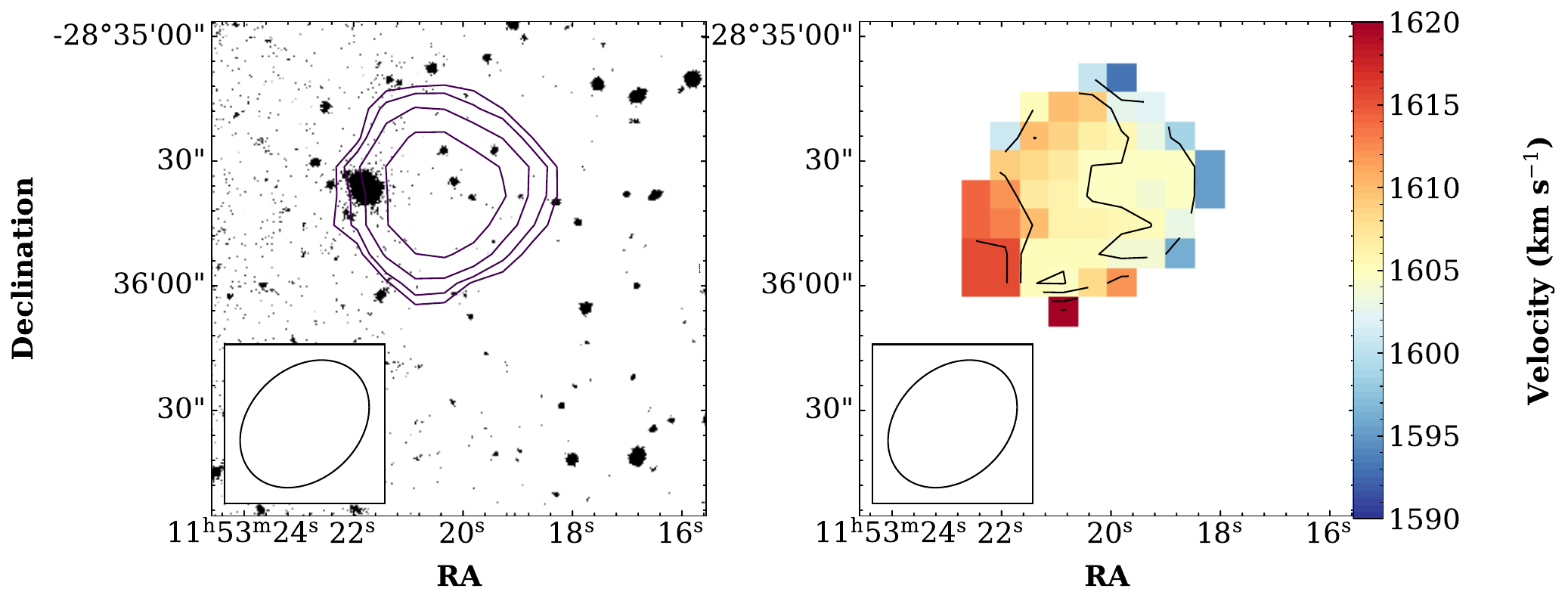}
    \caption{Left: The \HI intensity map (moment 0) and velocity field (moment 1) of the cloud. The plotted \HI contours are 3.3$\times$10$^{18}$ cm$^{-2} \times $ 2$^{n}$, n=0, 1, 2, 3,...Right: The velocity field of the cloud.}
    \label{fig:cloud_mom}
\end{figure}

\section{Results}
\label{sec:results}
 The source finding  yields the \HI emission associated with a total of 10 galaxies (including UGCA\,250), within the field of view of $\sim$~1.5$^{\circ}$ which corresponds to a projected radius of $\sim$~535 kpc at the distance of the galaxy. Within this field of view also lies the elliptical galaxy NGC 3923, positioned 30$^{'}$ (or approximately 200 kpc in projection) from UGCA\,250 and sharing a similar redshift ($V_{\rm sys}$ $\sim$ 1739 \kms). Despite its proximity, no \HI emission was detected in NGC\,3923.  The 5-sigma \HI mass upper limit for this non-detection is $10^{6.4}$ M$_\odot$, after correcting for the primary beam attenuation,  assuming a velocity width of 250 \kms and the same distance as that of UGC\,250.
 The \HI properties of the detected galaxies are given in Table \ref{table:satellites}. All the detected galaxies, except 6dFJ1151533-281047 are within 200 \kms of the systemic velocity of UGCA\,250.  The left panel of Fig. \ref{fig:mom0_overlay} shows the \HI contours of UGCA 250 and its satellite galaxies overlaid on  the VEGAS g--band optical image (The moment 0 and moment 1 maps of individual satellite galaxies are shown in Figs. \ref{fig:sat1} and \ref{fig:sat2}). The satellite galaxy IDs that are labeled are based on Table \ref{table:satellites}.
 The \HI contours are displayed at different resolutions: 26.7$\arcsec$ $\times$ 18.8$\arcsec$ ({\tt r10t00}) in purple, 34.3$\arcsec$ $\times$ 25.6$\arcsec$ ({\tt r15t00}) in cyan, 65.3$\arcsec$ $\times$ 63.8$\arcsec$ ({\tt r05t60}) in magenta, and 93.8$\arcsec$ $\times$ 91.7$\arcsec$ ({\tt r10t90}) in blue. The lower-resolution contours (magenta and blue) reveal both intra-group gas and gas around specific satellites such as ESO440--G023 (ID 1), 2MFGC09308  (ID 2), LEDA737259 (ID 3), and MKT115222-280126 (ID 8), and 6dFJ1151533-281047 (ID 10).

 We calculate the integrated flux of each detection using the {\tt SoFiA-2} output. Further, we calculate their \HI masses using the standard formula,  M$_{\rm HI} = 2.356\times 10^{5} D^{2} S_{\rm int} \ {\rm M}_{\rm \odot}$, where $D$ is the distance in Mpc, $S_{\rm int}$ is the integrated flux density in Jy~km~s$^{-1}$ \citep{Roberts62}. Six of the nine satellite galaxies (marked with $\ast$ in Table \ref{table:satellites}) were detected in the 5.5-hour single--track observations \citep{deBlok24}. Full-depth observations revealed three additional galaxies: 2MFGC 09308, MKT115222-280126, and MKT115551-282149, with \HI masses log(M$_{\rm HI}$/ M$_{\odot}$)  $\lesssim$ 6.60, consistent with the expected sensitivity limits of single--track and full-depth observations. MKT115551-282149, with log(M$_{\rm HI}$/ M$_{\odot}$) $\lesssim$ 6.10, was detected at a distance of 20 Mpc, demonstrating the remarkable sensitivity of the full-depth MHONGOOSE observations. Similarly, a low-mass, gas-rich, low-surface brightness galaxy in the Dorado Group was detected at 17.7 Mpc \citep{Maccagni24}. These results highlight the enhanced capabilities of full-depth observations in detecting low-mass galaxies at larger distances.

 
 \subsection{H{\sc I} Morphology of UGCA\,250}
 \label{sec:HI_morphology}

 The right panel of Fig. \ref{fig:mom0_overlay} displays a zoomed-in view of UGCA\,250 at a resolution of 26.7$\arcsec$ $\times$ 18.8$\arcsec$({\tt r10t00}),  which clearly shows the extent of gas at different column densities. Additionally, we show low-resolution contours in cyan, magenta, and blue, corresponding to 34.3$\arcsec$ $\times$ 25.6$\arcsec$, 65.3$\arcsec$ $\times$ 63.8$\arcsec$, and 93.8$\arcsec$ $\times$ 91.7$\arcsec$ respectively, in the outer regions. The red contour corresponds to a \HI column density of 1.0 $\times$ 10$^{20}$ cm$^{-2}$. Even at this column density, a noticeable position angle warp in the West direction is apparent. As we transition to lower \HI column densities, the disc does not expand significantly in the radial direction,
 but it does grow substantially in the vertical direction. The warp becomes more prominent at these lower column densities, and the \HI distribution of UGCA\,250 appears asymmetric, and slightly lopsided, with the disc column densities dropping more rapidly towards the northeast and southeast compared to the northwest and southwest. The thickening observed in the \HI disc is also apparent in the optical stellar disc (see Fig \ref{fig:optical}). 
 
 The galaxy has a diameter of $\sim$ 69 kpc at the projected \HI column density sensitivity of $\sim$ 3.3$\times$10$^{18}$ cm$^{-2}$. The prominent feature of this gas disc is that there is a tail like feature (henceforth filament) extending away from it in the southern direction up to $\sim$34 kpc (in projection).  The blob below this is at a projected distance of $\sim$ 41.5 kpc from the centre of galaxy, and has a \HI mass, log(M$_{\rm HI}$)~$\sim$~6.34. The \HI mass of filament along with the blob is, log(M$_{\rm HI}$)~$\sim$~7.01.

The moment maps derived at a slightly coarser resolution of 34.3$\arcsec$ $\times$ 25.6$\arcsec$ ({\tt r15t00}, $r =$ 1.5 and $t =$ 0) are shown in Fig. \ref{fig:UGCA250_moment_maps}. The left, middle, and right panels show the moment 0, moment 1, and moment 2 maps respectively. 
The middle panel shows the intensity weighted velocity field, which is rather complex. The inner disc is regularly rotating and the iso-velocity contours seem to turn over at lower \HI column densities. There is a cloud-like feature in the velocity field in the south of the galaxy which is at anomalous velocities. The visual inspection of the cube has revealed a cloud of gas coinciding with the main emission of the galaxy, but at a slightly different velocity, i.e at $\sim$ 1605 \kms\ in comparison to the systemic velocity $\sim$ 1694 $\kms \,.$
This can be seen in the channel maps presented in Fig. \ref{fig:channel_maps} in the channels with velocities from 1587 \kms\ to 1621 $\kms \,.$

The right panel shows the second moment map, indicative of velocity dispersion for symmetrical, single Gaussian profiles. We can see two very high velocity dispersion regions with values higher than 40 \kms\ . The high values near the cloud in the southern direction are not caused by intrinsically broader \HI profiles, but rather due to the cloud being at a different velocity compared to the galaxy's rotation. We separate this cloud and show its moment maps in Figure \ref{fig:cloud_mom}. 
The velocity field of the cloud shows a possible hint of rotation, but the velocity width is less than 15 \kms. Despite its \HI mass, log(M$_{\rm HI}$/ M$_{\odot}$) $\sim$ 6.80, we did not find any optical counterpart for this  cloud. Using the cloud size of 57$^{''}$ $\times$ 48$^{''}$ and assuming a stellar mass-to-light ratio of 1, we estimate a 5$\sigma$ upper limit for the stellar mass to be log(M$_{\ast}$/M$_{\odot}$) $\sim$ 6.1. There is also high velocity dispersion in the northeast of UGCA\,250. In the center, an x-shaped region with high dispersion values is observed (see right panel of Fig \ref{fig:UGCA250_moment_maps}), likely due to beam smearing associated with the rise of rotation curve and the inclination of this nearly edge-on galaxy \citep[e.g.][]{Voigtlander13}.

The features such as asymmetries, filament, cloud, and high velocity dispersion are also noticeable in the position-velocity (PV) diagrams along both the major and minor axes, shown in the appendix (Figs. \ref{fig:PV_min_1_2} and \ref{fig:lv_maps}). Hereafter, we refer to the major axis pv diagram as the lv diagram and the minor axis pv diagram as the bv diagram. The bv diagrams, taken at 30$\arcsec$ intervals along the minor axis, highlight features such as the cloud and the filament extending from the disk. The lv diagrams along the major axis indicate that the galaxy is more extended on the receding half, emphasizing an asymmetry that becomes more pronounced at higher vertical distances above the galactic plane.

\begin{figure}
    \centering
    \includegraphics[width=0.75\linewidth]{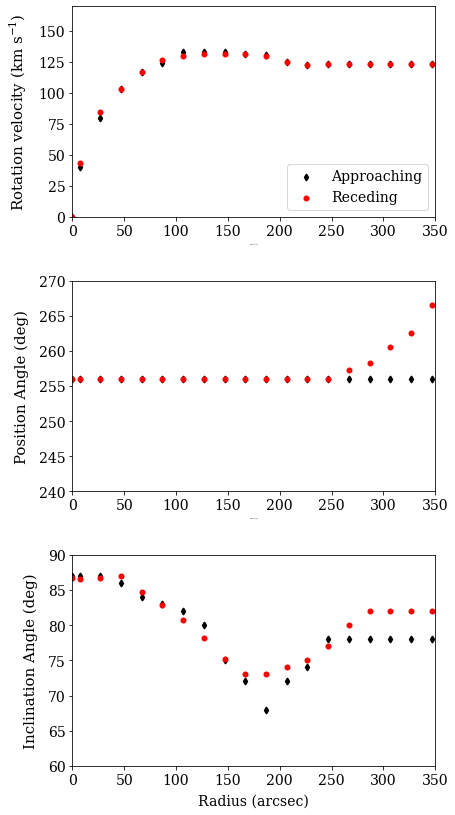}
    \caption{Top: de-projected rotation velocity, middle: position angle, bottom: inclination angle as a function of radius. }
    \label{fig:parameters}
\end{figure}

\begin{figure}
    \centering
    \includegraphics[width=0.78\linewidth]{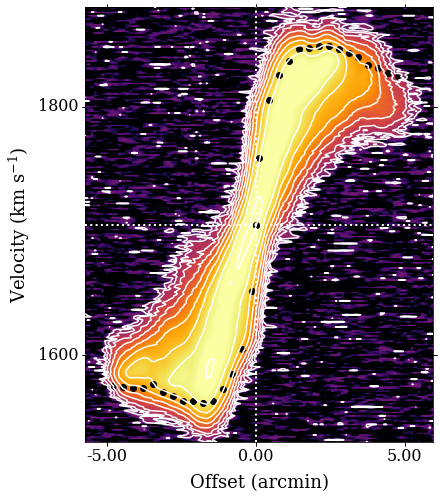}
    \caption{Position velocity diagram at the resolution of 34.3$\arcsec$ $\times$ 25.6$\arcsec$ ({\tt r15t00}, $r =$ 1.5 and $t =$ 0) along the major axis of UGCA\,250 (PA = 256$^{\circ}$ , V$_{sys}$ = 1705 \kms ). Contours begin at 3$\sigma$ and increase by factors of 2. The black circles represent the projected rotation velocities.}
    \label{fig:pv_maj_vrot}
\end{figure}


\subsection{HI Kinematics}
To investigate the morphology and kinematics of the \HI in this galaxy, we construct model data cubes and compare them to the observations. We use the Tilted Ring Fitting Code (TiRiFiC) software \citep{Jozsa07} for modelling. This software uses a tilted ring approach, representing a galaxy as concentric rings each characterized by specific kinematic and orientation parameters \citep{Rogstad74}. These parameters, including central position, systemic velocity, inclination angle with respect to line of sight, position angle of the receding major axis, rotational velocity, velocity dispersion, vertical thickness, and surface brightness distribution  can either be fixed or varied with radius. All the modelling described below uses the cubes at resolution of 34.3$\arcsec$ $\times$ 25.6$\arcsec$ ({\tt r15t00}, $r =$ 1.5 and $t =$ 0). The approaching and receding halves of UGCA\,250 are modelled separately since the galaxy clearly shows asymmetries.
We first use Fully Automated TIRIFIC (FAT) software \citep{Kamphuis15}, which is a wrapper around TiRiFiC, to derive the initial guess model, i.e., the initial de-projected \HI surface brightness profile and the de-projected rotation curve of the thin disc. Then, we use these parameters as an initial guess for the data cube using TiRiFiC software. An initial estimate for the systemic velocity, center, position angle, and inclination angle were obtained from the FAT run. These matched well with the rough estimates of kinematic center and systemic velocity found using the zeroth and first moment maps. The radial variation of position angle and inclination angle were then derived while the galaxy center and systemic velocity were fixed at the values found above. The position angle (PA) seems to show some radial variation; we find that the value of PA changes only in the outer regions of the receding half and stays constant in the approaching half, as can also be seen in the channel maps of the galaxy in Fig. \ref{fig:channel_maps} in the velocities ranging from 1805 to 1838 \kms. The radial variation of PA was adjusted based on channel maps and the total \HI map. We show the PA and inclination angle as a function of radius in the middle and bottom panels of Fig. \ref{fig:parameters}. The radial variation of inclinations suggests the \HI disc to be warped along the line of sight, as seen in the moment 0 map. All of these parameters were later refined both by comparing observed and modeled position–velocity diagrams, channel maps, and zeroth-moment maps. The rotation curve was adjusted based on the major-axis position-velocity (pv) diagrams tracing emission along the edge with the highest velocities. Figure \ref{fig:pv_maj_vrot} shows a position-velocity diagram along the major axis with the derived rotation curve (projected) overlaid on it. The parameters, systemic velocity (1705 \kms), kinematic center (11$^{h}$ 53$^{m}$ 24.4$^{s}$, -28$^{\circ}$ 33$^{'}$ 9.5$^{''}$) and the run of the position angle and inclination angle, column density profile, and rotation curve are kept constant throughout all of the models for the entire process. This model, which includes the PA and inclination angle warp component fits the thin disc well, is considered the base model for the galaxy. In the following section, we will include additional components to our model in order to reproduce the HI emission from the EPG.

\begin{figure*}
    \centering
    \includegraphics[width=0.80\linewidth]{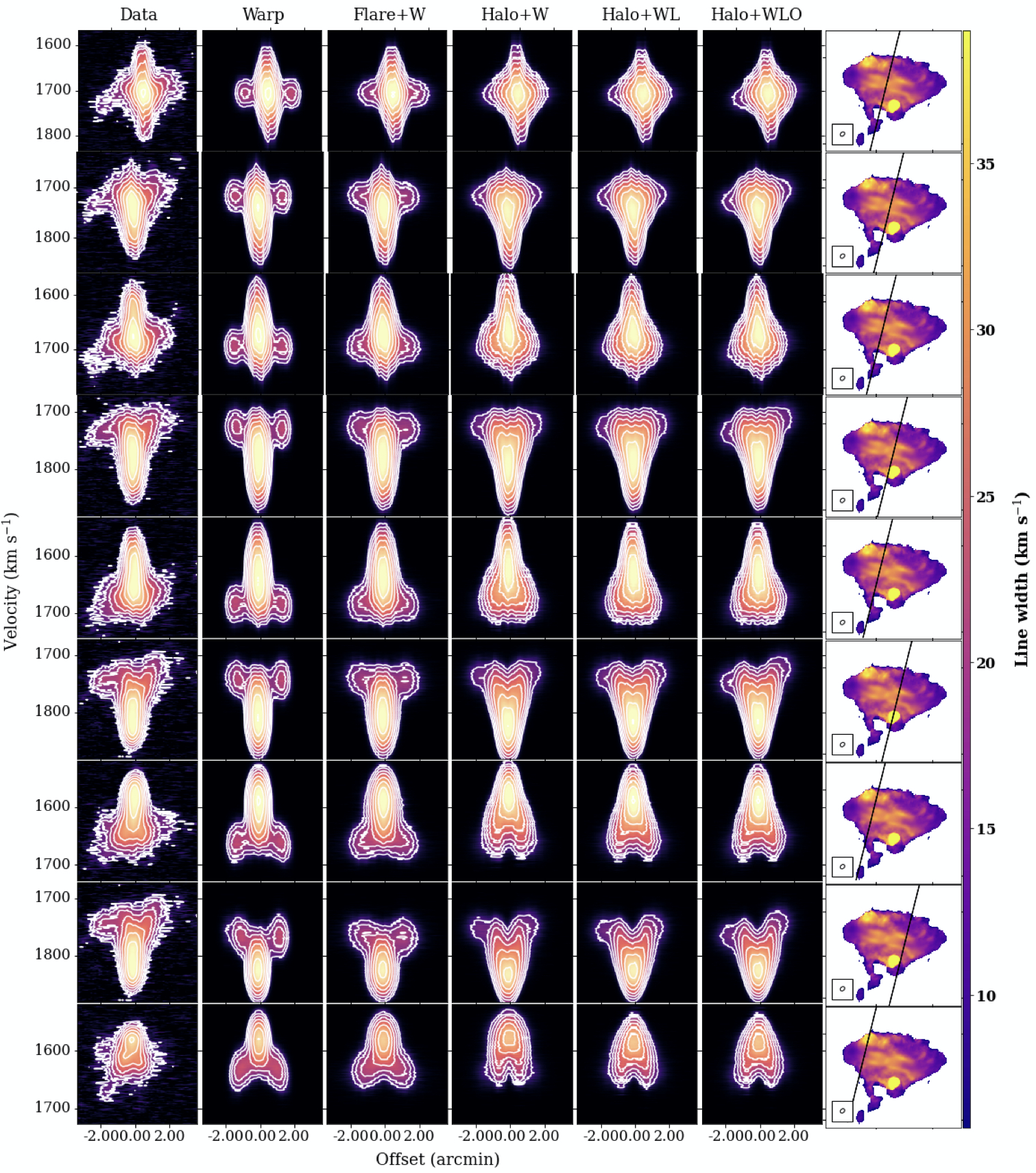}
     \caption{Minor axis position–velocity diagrams of UGCA\,250 and models. Slice locations are shown in the left panel of Fig. \ref{fig:UGCA250_moment_maps}. Contours begin at 3$\sigma$ and increase by factors of 2. All the models include a warp across the line of sight. From the left: the data at 34.3$\arcsec$ $\times$ 25.6$\arcsec$ resolution (Data); one component model made of a thin disc (scale height $\sim$ 200 pc) with a warp (Warp); a flare with a warp (Flare+W); two component model made of a thin disc and a corotating halo (scale height $\sim$ 3kpc) with a warp (Halo+W); a two component model made of a thin disc and a lagging halo and warp  (Halo+WL); two component model with a warp, lagging halo, and a radial motion outwards in the outer regions of galaxy (Halo+WLO). The rightmost column displays the moment 2 map, with the line in each panel indicating the slice location corresponding to the bv diagram in the respective row.    }
     \label{fig:pv_models}
\end{figure*}

\begin{figure*}
    \centering
    \includegraphics[width=0.85\linewidth]{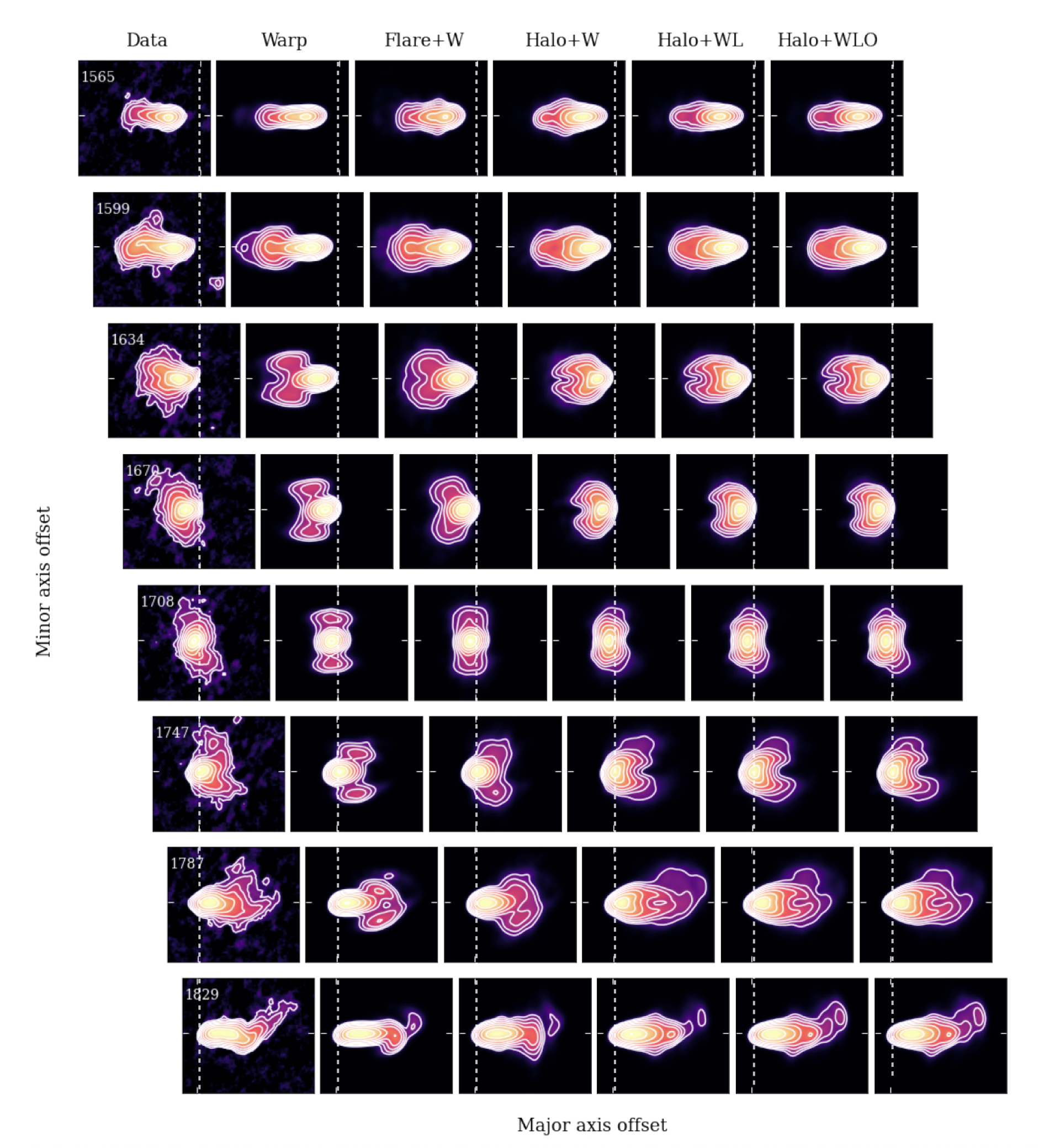}
     \caption{Channel map comparison of UGCA\,250 observations with various models. The heliocentric velocities are shown on the top left of data frames. All channel maps are aligned such that their centers match, with a white dashed line drawn across the center for all maps.  From the left: the data at 34.3$\arcsec$ $\times$ 25.6$\arcsec$ resolution (Data); one component model made of a thin disc (scale height $\sim$ 200 pc) with a warp (Warp); a flare with a warp (Flare+W); two component model made of a thin disc and a corotating halo (scale height $\sim$ 3kpc) with a warp (Halo+W); a two component model made of a thin disc and a lagging halo and warp  (Halo+WL); two component model with a warp, lagging halo, and a radial motion outwards in the outer regions of galaxy (Halo+WLO).  }
     \label{fig:channel_models}
\end{figure*}

\subsection{Models}
In order to understand if the anomalous gas has truly anomalous kinematics or if it is produced by projection effects due to inclination angle or thickness of the disc, etc., we develop several models. We start with the simplest possible model and then add various features (e.g., a warp or flare) one at a time to understand the subtle distinctions between various features. We visually compare the resulting model with data, i.e., we examine the bv position-velocity diagrams (i.e., along the minor axis) as well as channel maps of models and data to fine tune parameters in models. We consider several models, including a simple one-component which is a single disc with a vertical sech$^{2}$ profile, a flare where the scale height varies as a function of radius, and two components with a thin disc and a thick corotating disc. Finally, we consider a more sophisticated model of a thin disc and a thick disc with different kinematics and we explore the effect of radial motions (i.e., inflows or outflows). All models include a PA and inclination angle warp component. We have verified that all models that do not include a warp fail at reproducing several of the basic features observed in the data, and will not be considered here.

The selected channel maps shown in Fig. \ref{fig:channel_maps} and the bv diagrams shown in Fig. \ref{fig:PV_min_1_2} display a number of distinct features sensitive to various model parameters, which were used to distinguish between different models. In the bv diagrams, we can clearly see the anomalous cloud and the filament features of the galaxy. This cloud and tidal tail features were not included in the model fitting, as they are distinct features not described by the thin or thick disc models. Some of the prominent features that were considered while modeling were the spacing and orientation of the contours. Because of the high number of degrees of freedom, the parameters of the best-fit model from TiRiFiC tend to have strong discontinuities, with abrupt changes in fitted values across the parameter space. Furthermore, TiRiFiC employs $\chi^2$ minimization, which inherently gives higher weight to regions with high signal-to-noise ratios. However, it is worth noting that these regions are not necessarily the most suited to distinguish between models. Therefore, while TiRiFiC's automated fitting capabilities were used when appropriate, manual intervention becomes necessary to guide the fitting process effectively \citep[e.g.][]{Oosterloo07, Zschaechner12, Kamphuis13}.

To determine whether the observed thickness of the \HI disc can be attributed to a warp in the thin disc, we initially consider a simple model consisting of a single thin disc warped approximately 15$^{\circ}$ along the line of sight, which is our base model. Figures \ref{fig:pv_models} and \ref{fig:channel_models} show the bv diagrams and channel maps of both the observed data and the various models respectively. The observed data is presented in the leftmost column, while the second column shows the model of a thin disc with a scale height $\sim$ 200 pc, comparable to that observed in the Milky Way and other similar edge-on galaxies. The morphology of the channel maps and bv diagrams, particularly near the systemic velocity, in this model is markedly different from the observed data, emphasizing that the model fails to accurately reproduce the observed features. In addition, we find that no amount of disc warping or disc thickness alone could reproduce the data adequately. We also tested increasing the velocity dispersion, which tended to thicken the emission from the channel maps in the vertical direction, but this change did not consistently match the morphologies observed in the channel maps.

We then add a substantial flare to the warped disc model (or base model), denoted as `Flare+W' in the figures, by increasing the scale height with radius, which allows for a better fit to the bv diagrams and channel maps. The scale height increases from 0.2 kpc in the inner regions to 5 kpc in the outer regions. This modified model shows an improvement compared to the previous thin disc+warp model. The emission on the systemic sides of the bv diagrams and the thickening in the channel maps are now more accurately reproduced. The improvements due to the flare are most evident on the systemic velocity sides of the bv diagrams. However, shape,  spacing, and orientation of the outer contours near the systemic side still show differences compared to the observed data.

We further explore a two-component model, consisting of a thin disc and a thick disc/halo, denoted as `Halo+W' in the figures. \HI  haloes have been observed in several galaxies, with NGC 891 being among the most notable \citep{Oosterloo07}.  For the thin disc, we assume a scale height of 0.2 kpc with a vertical sech$^{2}$ profile. The deprojected surface density profile of the thick disc is taken to be a scaled version of the thin disc profile. We test different scale heights for the thick disc. The morphology of bv diagrams depends on the scale-height of the thick disc and  varying the scale heights between 2 kpc and 4 kpc give broadly similar results,  so we use a scale height of $\sim$ 3  kpc for the thick disc. This is similar for NGC 3198 as presented in \citet{Gentile13}. This model significantly improves the reproduction of the thickening in the minor axis direction evident in the channel maps at systemic velocities. It also aligns well with the observed data, particularly the spacing and orientation of the outer contours near the systemic velocity side in bv diagrams.

However, the channel maps at the receding side velocities still do not match well with this model, which assumes that the halo is corotating with the thin disc, i.e., there is no lag. To address this, we introduced a lag that varies with radial distance from the center, now labeled as `Halo+WL' in the figures. This lag causes the observed velocity of the gas to be shifted in the direction of the systemic velocity, with this shift increasing with height. While the bv diagrams of the no lag and lag models look very similar, we were mainly able to distinguish these models using the channel maps. The fine-tuned model shows a much better match in the channel maps at terminal side velocities, especially on the receding side (see last row of Fig \ref{fig:channel_models}). The lag in these central regions is about 20 \kms kpc$^{-1}$ starting at radius, R $\sim$ 4.6 kpc, remaining constant out to R $\sim$ 6.6 kpc, then decreasing to zero at R $\sim$ 20 kpc in the outer regions on the approaching side. Conversely, on the receding side, the lag is about 20 \kms kpc$^{-1}$ starting at R $\sim$ 2.6 kpc, remaining constant out to R $\sim$ 8.6 kpc, then decreasing to zero at R $\sim$ 16.5 kpc in the outer regions.

Additionally, while models with uniform lag throughout were considered, the fine-tuned model with radial variation proved to provide a better description of the data.  Finally, we explored a model incorporating radial motions of the anomalous gas in the outer regions, indicated as `Halo+WLO' in the figures. This model is similar to the fine-tuned model, but it includes a radial motion outward from the galaxy at 10 \kms\ in the outer regions. This adjustment accurately reproduces the direction of the outermost contour in the receding half of the bv diagrams, demonstrating its effectiveness in modeling specific features of the galaxy. To determine the orientation of the disk, we used the optical image, which shows distinct dust lanes offset to the North; this offset indicates that the northern rim of the galaxy is the near side. This is crucial, as the interpretation of radial motions, whether characterized as outflows or inflows, relies on accurately identifying the near and far sides \citep{Marasco19}. 

\begin{figure}
    \centering
    \includegraphics[width=0.99\linewidth]{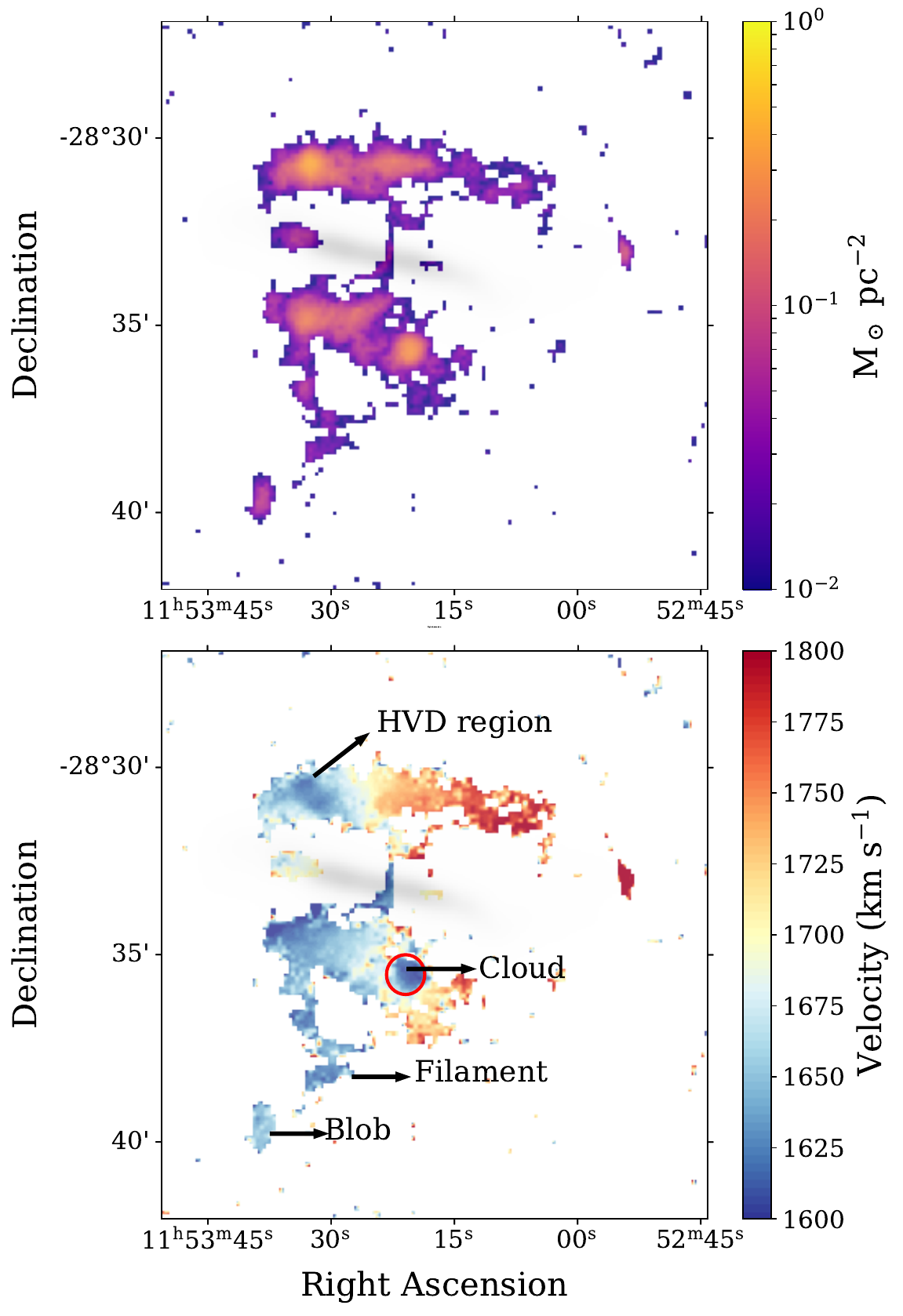}
    \caption{Top: \HI gas surface density of the anomalous gas that was not modelled. Bottom: the velocity field of the anomalous gas , the colour shows the velocity in \kms. }
    \label{fig:anomalous}
\end{figure}

\begin{figure*}
    \centering
    \includegraphics[width=0.49\linewidth]{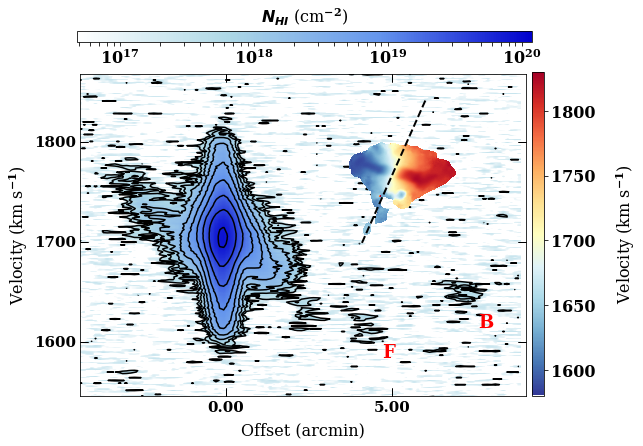}
    \includegraphics[width=0.49\linewidth]{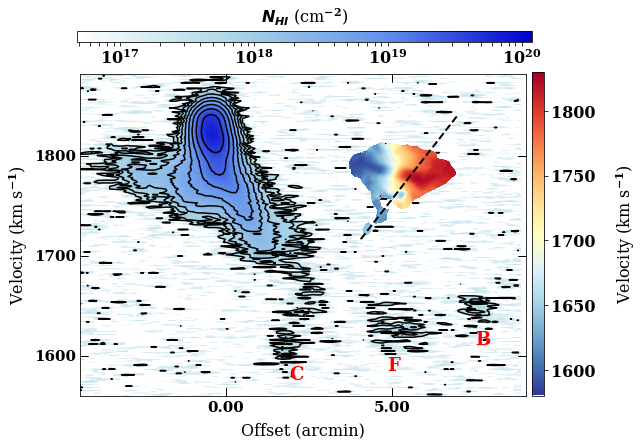} 
    \includegraphics[width=0.49\linewidth]{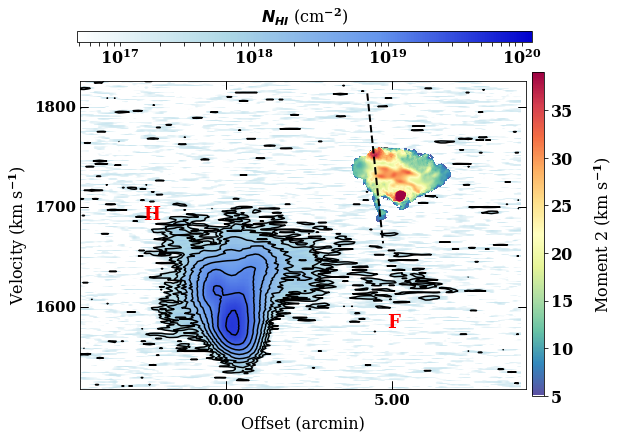}
    \includegraphics[width=0.49\linewidth]{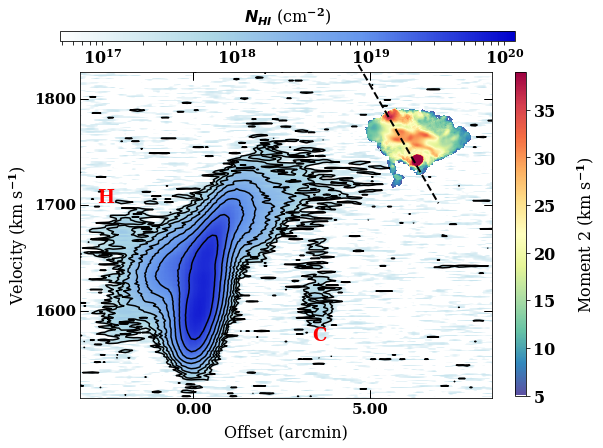}
    \caption{ Top left: Position-velocity (PV) diagram through the filament region of the galaxy. The inset in the top right corner shows the moment 1 map with a black dashed line indicating the PV slice path. Top right: PV diagram through the filament and cloud regions. The inset also displays the moment 1 map with the black dashed path. Bottom left: PV diagram across the high velocity dispersion region and filament. The inset shows the moment 2 map with the black dashed path. Bottom right: PV diagram across the high velocity dispersion region and the cloud. The inset again displays the moment 2 map with the same black dashed line. The horizontal color bar represents \HI column density for the PV diagrams, while the vertical color bar represents either velocity (moment 1) or velocity dispersion (moment 2) in \kms. In the maps, C denotes the cloud at anomalous velocities, F indicates the filament, B refers to the blob below the filament, and H represents high velocity dispersion. }
    \label{fig:PVD}
\end{figure*}
\section{Discussion}
\label{sec:disc}

The key question of this work concerns the origin of the extraplanar and anomalous gas in UGCA\,250. 
Specifically, the extraplanar gas has been proposed to originate either from processes within the galactic disc, such as superbubbles generated by supernovae and stellar winds, or from external mechanisms, including interaction with nearby companions or the accretion of primordial gas from the cosmic web \citep{Fraternali06, Sanchez14, Marasco19}
Understanding the origin of this extraplanar and anomalous gas is crucial for unraveling the processes governing the dynamics and evolution of galaxies. 

Our detailed modelling of UGCA\,250 suggests that approximately $15 \pm 5\%$ of the galaxy's \HI content is distributed within a diffuse halo with  a scale height of 3 $\pm$ 1 kpc. This result is consistent with \citet{Marasco19}, which shows that extraplanar \HI is nearly ubiquitous in star forming disc galaxies, typically making up about about 5-25$\%$ of the total \HI mass, with a mean value of 14$\%$, and a typical thickness of a few kpc.  A noticeable warp is also evident, not only seen in the zeroth-moment map, but is also captured in our extensive modeling. The fine-tuned model, which incorporates a radially varying lag —approximately $20$ km s$^{-1}$ in the central regions that tapers to zero at R $\sim$ 20 kpc— appears to best fit the data. We found no substantial evidence supporting the accretion of primordial gas, as there are no clear kinematical signatures of inflowing gas in our data. However, given the complex dynamics arising from tidal interactions and the presence of nine satellite galaxies, detecting such signatures, even if they exist, would be challenging. The most definitive way to identify primordial gas accretion is through metallicity measurements, as they can reveal the low metallicities, characteristic of such gas. Unfortunately, these measurements are not available for this study, limiting our ability to conclusively determine the presence or absence of primordial gas.
Additionally, there is some evidence of radial motions in the outer regions of the receding side, with an outward direction. Models using outflows only in the central regions, however, could not reproduce the data accurately, particularly the outermost contour in the bv diagrams (as seen in Fig \ref{fig:pv_models}). This indicates these outflows might not result from star formation. Instead, they could be linked to radial motions triggered by potential tidal interactions, which is further corroborated by the detection of 9 satellite galaxies and intra-group gas.  

Despite extensive modeling, there remains a significant amount of \HI that is spatially or kinematically distinct from the disc gas. To isolate this emission, we blank the data where the model values exceed $0.5 \sigma$, where $\sigma$ is the noise in the observed cube. 
This results in a ``residual'' data cube from which we can generate the moment maps of the anomalous gas, i.e. the gas not described by either the thin or thick disc models. These maps are presented in Figure \ref{fig:anomalous}. The total \HI mass of this anomalous gas is approximately $10^{7.7} \text{ M}_\odot$, which accounts for about 0.7$\%$ of the total mass of the galaxy. Although this is a relatively small fraction and may not significantly impact the overall \HI content, it could be tracing underlying ionized gas and is therefore relevant. We observe that the thin disc and most of the thick disc are well modeled. However, three prominent features of the anomalous gas are identified: (i) The primary feature is a tail-like filament extending away from the disc in the southern direction up to approximately 34 kpc (in projection). Below this filament is a blob situated at a projected distance of about 41.5 kpc from the center of the galaxy, with an \HI mass of  log(M$_{\rm HI}$)~$\sim$~6.34. The combined \HI mass of the filament and this blob is log(M$_{\rm HI}$)~$\sim$~7.01. It is worth noting that satellite ID 3 (see Fig \ref{fig:mom0_overlay})  is located below the filament and blob, at a distance of 65 kpc from the blob. (ii) Further,  a cloud-like feature in the velocity field south of the galaxy exhibits anomalous velocities. Visual inspection of the cube reveals a cloud of gas coinciding with the main emission of the galaxy but at a different velocity, visible in the channel maps in Figure \ref{fig:channel_maps}. No optical counterpart for this cloud has been detected. (iii)  There is a region with high velocity dispersion located in the northeast of the galaxy, on the approaching side.

In Figure \ref{fig:PVD}, the position-velocity (PV) diagrams provide further insights into these features. The top left panel shows a PV diagram through the filament region of the galaxy, while the top right panel depicts a PV diagram through the filament and cloud regions. The bottom right panel presents a PV diagram across the high-velocity dispersion region and the cloud, and the bottom left panel shows a PV diagram across the high-velocity dispersion region and the filament. Each panel contains an inset indicating the path of the PV slice on the corresponding moment maps. These PV diagrams through these anomalous features hint at a history of close encounters or mergers, leading to tidal disturbances that generated filament-like structures and cloud-like anomalies. The interconnected patterns across the cloud and filament regions support the hypothesis of a prior interaction. 

It is important to note that these various features including the filament, cloud, and high velocity dispersion region were not detected with the 5.5-hour track of MHONGOOSE observations \citep{deBlok24}, which have a 3$\sigma$ column density sensitivity of approximately $6.7 \times 10^{19}$ atoms per cm$^{-2}$. As illustrated in Fig. \ref{fig:mom0_overlay} , at this column density sensitivity, the vertical distribution of the gas or thick disc is also less apparent. The full 55 hours of MeerKAT data, however, reveal these features in significant detail, highlighting the advantages of its superior column density sensitivity. This has led to the detection of 9 satellite galaxies, with six identified in the single track observations and three additional galaxies revealed in the full depth data.
Furthermore, at lower resolutions and column densities, intra-group gas was also detected along with the satellites. Additionally, the shell elliptical galaxy NGC 3923, which is located about 30 arcminutes (or 200 kpc) from UGCA\,250, falls within our observation's field of view and has a redshift similar to UGCA\,250's. This galaxy is known for its extensive system of stellar shells, the most of any shell galaxy, which are thought to be remnants of past galactic interactions \citep[e.g.][]{Bilek16, Spavone17}. 

The zoomed-in moment 0 and moment 1 maps of the individual satellite galaxies are shown in the appendix figures \ref{fig:sat1} and \ref{fig:sat2}. In these maps, we observe \HI\ emission offset from the stellar counterpart in several cases (e.g., IDs 1, 4, 5, and 7), along with evident asymmetries and disturbed morphologies that indicate interactions within these systems. Further, all detected satellite galaxies lie within the 356 kpc field of view of MeerKAT, and, except for satellite ID 10, all are within 200 km/s of the systemic velocity of UGCA\, 250, Hence, UGCA\,250, NGC\,3923, and all the detected satellites likely belong to the same group, as all are contained within the group’s estimated virial radius of 591 kpc and virial mass of 10$^{13.3}$ M$_{\odot}$ \citep{Kourkchi17}.  This suggests that the larger environment of UGCA\,250, including its satellites and NGC\,3923, appears to be a dynamically active system.
The distinct anomalous gas features suggest tidal interactions or a past event involving something passing close to UGCA\,250 as the likely source of this gas. This interpretation is further supported by the optical discontinuity observed in UGCA 250 (see Fig. \ref{fig:optical}). The discontinuity is likely associated with the main disc rather than a separate dwarf galaxy, as no H{\sc i} emission at a distinct velocity is detected at that location. 
 Further, the observation of satellite galaxies and detection of intra-group gas indicate potential close encounters or mergers that could create tidal disturbances manifesting as filaments and clouds with peculiar velocities.
The outward radial motions in the outer receding regions further point to these interactions shaping the unique extraplanar gas kinematics and structure. 
 A more detailed study of the distribution and interactions of the satellite galaxies is beyond the scope of this paper.

While environmental effects likely play a key role in shaping the extraplanar gas in UGCA\,250, the contribution of stellar feedback must also be considered. UGCA\,250 is a regular star-forming galaxy on the main sequence, and to quantify the role of stellar feedback, we follow the framework of \citet{Marasco19}, applying their Equation 11 with a mass-loading factor $\beta = 7$ to estimate the expected extraplanar gas mass from a galactic fountain flow model. This yields log(M$_{\rm EPG}$ / M$_{\odot}$)  $\sim$ 8.6, about 0.4 dex lower than our observed value. This discrepancy suggests that, while stellar feedback likely accounts for a significant portion of the thick disc, tidal interactions are responsible for the more extended features, such as the filament and cloud. UGCA\,250 shares several characteristics with NGC\,2403, including rotational velocity, star formation rate, and extraplanar gas properties, where a galactic fountain model interacting with the hot circumgalactic medium accounts for much of the extraplanar gas \citep{Li23}. Additionally, NGC\,2403 is part of a group of five galaxies \citep{Kourkchi22}, with some of its extraplanar \HI attributed to remnant tidal interactions with a satellite \citep{Veronese23}. We therefore propose a mixed origin for the extraplanar gas in UGCA\,250, with the thick disc primarily shaped by stellar feedback and the outer structures influenced by tidal interactions. This study highlights the complex interplay between internal galactic processes and external environmental factors shaping the anomalous gaseous structures around UGCA\,250.

\section{Conclusions}
\label{sec:summary}

We use ultra-deep MHONGOOSE survey observations to investigate the extraplanar gas in UGCA\,250. The survey's combination of high column density sensitivity and high spatial resolution—ranging from 6.0$\times$10$^{19}$ cm$^{-2}$ at 6$''$ to 5.7$\times$10$^{17}$ cm$^{-2}$ at 90$''$ (3$\sigma$ over 16 \kms ). This offers an ideal framework for examining the physics of extraplanar gas and the disc-halo connection. This has enabled the detection of nine satellite galaxies, a distinctive filament extending approximately 41 kpc southward from the galaxy, and a cloud at anomalous velocities. We have not detected any counterpart for this cloud in the VST deep optical image.

We construct a detailed tilted ring model for this edge-on galaxy to gain a deeper understanding of the vertical structure of its neutral hydrogen. Through detailed modelling, we estimate that approximately 15 $\pm$ 5  $\%$  of the galaxy's \HI content resides in a halo extending 3$\pm$ 1 kpc above the galactic plane. This halo exhibits a lag that varies with radial distance from the center,  beginning at 20 km/s within the central regions and tapering to zero at a radial distance of 20 kpc. Additionally, we found evidence for radial motions in the outward directions in the outer regions of the receding half of the galaxy. The outermost contours in the bv diagrams can only be reproduced by incorporating outflows in the outer regions. Since there is little star formation occurring in these outer regions, these outflows are likely caused by tidal interactions rather than stellar feedback.
The combination of radial motions in the outer regions, multiple satellite galaxies, and distinct features such as the cloud, filament, and high-velocity dispersion region suggests that tidal interactions are at least in part responsible for the formation of anomalous gas in UGCA\,250. Therefore, our analysis suggests a mixed origin for the extraplanar gas in UGCA 250, with contributions from both internal processes, such as stellar feedback, and external influences, including tidal interactions with satellite galaxies.


\section*{Acknowledgements}

SK, DJP, NZ, and SHAR are supported by the South African Research Chairs Initiative of the Department of Science and Technology and National Research Foundation (Grant number 77825). This work has received funding from the European Research Council (ERC) under the European Union’s Horizon 2020 research and innovation programme (grant agreement No. 882793 ``MeerGas''). PK and RJD are partially supported by the BMBF project 05A23PC1 for D-MeerKAT. JH acknowledges support from the UK SKA Regional Centre (UKSRC). The UKSRC is a collaboration between the University of Cambridge, University of Edinburgh, Durham University, University of Hertfordshire, University of Manchester, University College London, and the UKRI Science and Technology Facilities Council (STFC) Scientific Computing at RAL. The UKSRC is supported by funding from the UKRI STFC. EA and AB acknowledge support from the Centre National d'Etudes Spatiales (CNES), France. L.C. acknowledges financial support from the Chilean Agencia Nacional de Investigación y Desarrollo (ANID) through Fondo Nacional de Desarrollo Científico y Tecnologico (FONDECYT) Regular Project 1210992. LVM acknowledges financial support from the grant PID2021-123930OB-C21 funded by MICIU/AEI/ 10.13039/501100011033 and by ERDF/EU and grant CEX2021-001131-S funded by MICIU/AEI.

 The MeerKAT telescope is operated by the South African Radio Astronomy Observatory, which is a facility of the National Research Foundation, an agency of the Department of Science and Innovation. We acknowledge the use of the ilifu cloud computing facility—www.ilifu.ac.za, a partnership between the University of Cape Town, the University of the Western Cape, the University of Stellenbosch, Sol Plaatje University, the Cape Peninsula University of Technology, and the South African Radio Astronomy Observatory. The Ilifu facility is supported by contributions from the Inter-University Institute for Data Intensive Astronomy (IDIA a partnership between the University of Cape Town, the University of Pretoria, the University of the Western Cape and the South African Radio Astronomy Observatory), the Computational Biology division at UCT and the Data Intensive Research Initiative of South Africa (DIRISA).

\section*{Data Availability}

The moment maps of MHONGOOSE observations of UGCA\,250 are available for download at the MHONGOOSE website \url{(https://mhongoose.astron.nl)}



\bibliographystyle{mnras}
\bibliography{UGCA250} 




{\noindent
\it
\small
\\
$^{1}$ Department of Astronomy, University of Cape Town, Private Bag X3, Rondebosch 7701, South Africa\\
$^{2}$Netherlands Institute for Radio Astronomy (ASTRON), Oude Hoogeveensedijk 4, 7991 PD Dwingeloo, the Netherlands\\
$^{3}$Kapteyn Astronomical Institute, University of Groningen, PO Box 800, 9700 AV Groningen, The Netherlands \\
$^{4}$Ruhr University Bochum, Faculty of Physics and Astronomy, Astronomical Institute (AIRUB), 44780 Bochum, Germany.\\
$^{5}$South African Astronomical Observatory (SAAO), PO Box 9, Observatory, 7935, Cape Town, South Africa \\ 
$^{6}$Jodrell Bank Centre for Astrophysics, School of Physics and Astronomy, University of Manchester, Oxford Road, Manchester M13 9PL, UK\\
$^{7}$ United Kingdom SKAO Regional Centre (UKSRC), UK\\
$^{8}$INAF – Osservatorio Astronomico di Cagliari, via della Scienza 5, 09047, Selargius (CA), Italy\\
$^{9}$Aix Marseille Univ, CNRS, CNES, LAM, Marseille, France \\
$^{10}$Argelander-Institut für Astronomie, Auf dem Hügel 71, 53121, Bonn, Germany \\
$^{11}$Centre for Astrophysics Research, University of Hertfordshire, College Lane, Hatfield, AL10 9AB, UK \\
$^{12}$Instituto de Astrofísica, Departamento de Ciencias Físicas, Universidad Andrés Bello, Fernandez Concha 700, Las Condes, Santiago, Chile \\
$^{13}$Universit$\acute{e}$ de Strasbourg, CNRS, Observatoire astronomique de Strasbourg, UMR 7550, 67000 Strasbourg, France
$^{14}$Observatoire de Paris, Collège de France, Université PSL, Sorbonne Université, CNRS, LERMA, Paris, France \\
$^{15}$Max-Planck-Institut für Radioastronomie, Auf dem Hügel 69, 53121, Bonn, Germany \\
$^{16}$Department of Physics and Electronics, Rhodes University, PO Box 94, Makhanda, 6140, South Africa \\
$^{17}$Australia Telescope National Facility, CSIRO, Space and Astronomy, PO Box 76, Epping, NSW, 1710, Australia \\
$^{18}$School of Science, Western Sydney University, Locked Bag 1797, Penrith, NSW 2751, Australia \\
$^{19}$INAF - Padova Astronomical Observatory, Vicolo dell'Osservatorio 5, 35122, Padova, Italy \\
$^{20}$International Centre for Radio Astronomy Research, The University of Western Australia, 35 Stirling Highway, Crawley, WA, 6009, Australia \\
$^{21}$Southern African Larger Telescope (SALT), PO Box 9, Observatory, 7935, Cape Town, South Africa \\
$^{22}$Max-Planck-Institut für Astronomie, Königstuhl 17, D-69117, Heidelberg, Germany\\
$^{23}$Instituto de Astrofísica de Andalucía-CSIC, Glorieta de la Astronomía s/n, 18008, Granada, Spain \\
$^{24}$Department of Physics and Space Science, Royal Military College of Canada, PO Box 17000, Station Forces Kingston, ON, K7K 7B4, Canada \\
$^{25}$Department of Physics, Engineering Physics and Astronomy, Queen's University, Kingston, ON, K7L 3N6, Canada \\}

\appendix

\section{PV diagrams of UGCA250}
\begin{figure*}
    \centering
    \includegraphics[width=0.49\linewidth]{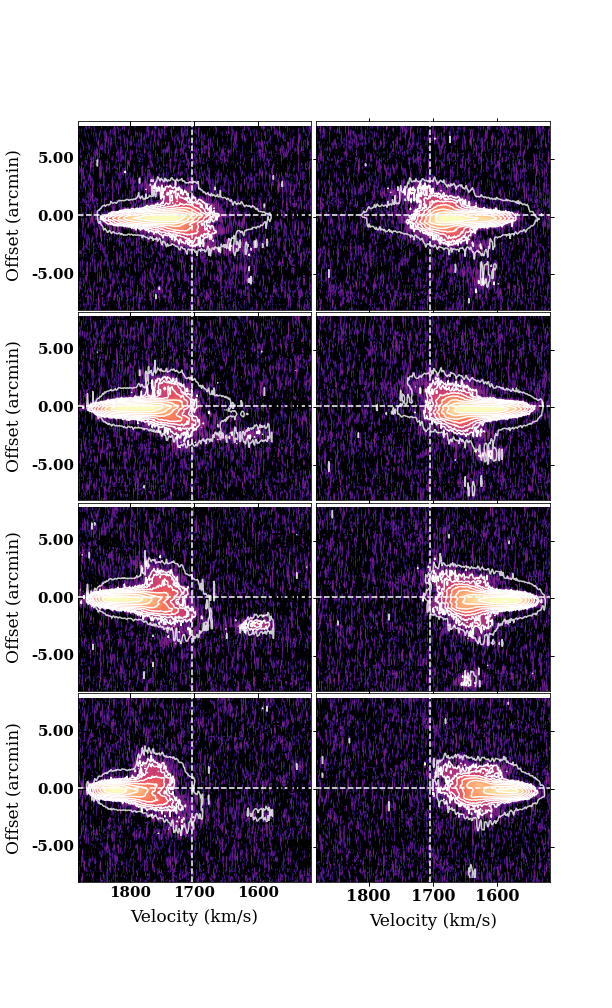}
    \includegraphics[width=0.49\linewidth]{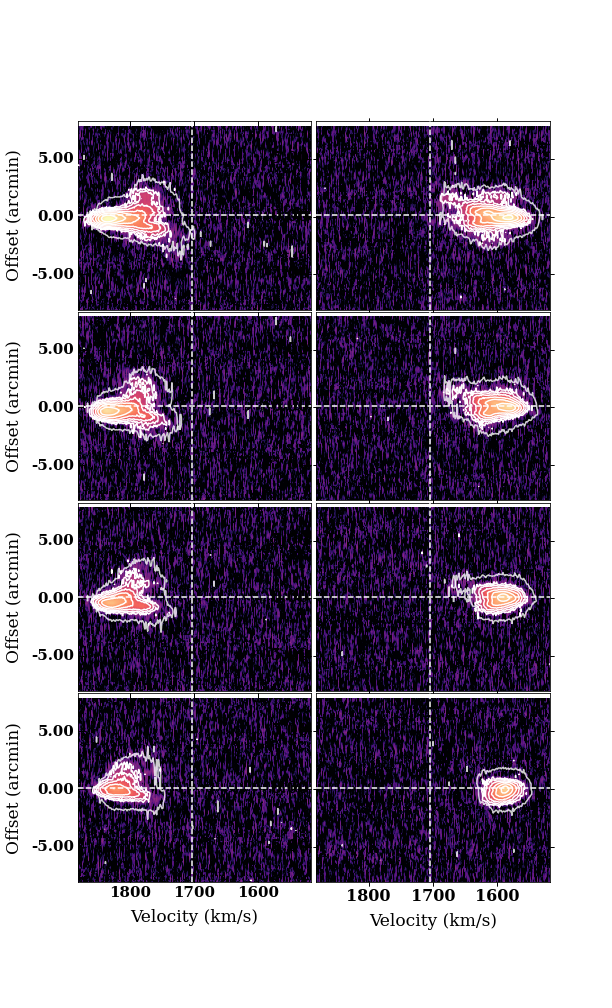}
    \caption{The bv diagrams of UGCA\,250 at the resolution of 34.3$\arcsec$ $\times$ 25.6$\arcsec$ ({\tt r15t00}, $r =$ 1.5 and $t =$ 0).  The diagrams correspond to slices taken at offsets of 30, 60, 90, 120, 150, 180, 210, and 240 arcseconds from the centre, separately for the approaching and receding halves of the galaxy. Contours begin at 3$\sigma$ and increase by
factors of 2.  The outermost contour of each map is the 3$\sigma$ contour at the resolution of 65.3$\arcsec$ $\times$  63.8$\arcsec$ ({\tt r05t60}, $r =$ 0.5 and $t =$ 60$\arcsec$).}
    \label{fig:PV_min_1_2}
\end{figure*}

\begin{figure*}
    \centering
    \includegraphics[width=0.78\linewidth]{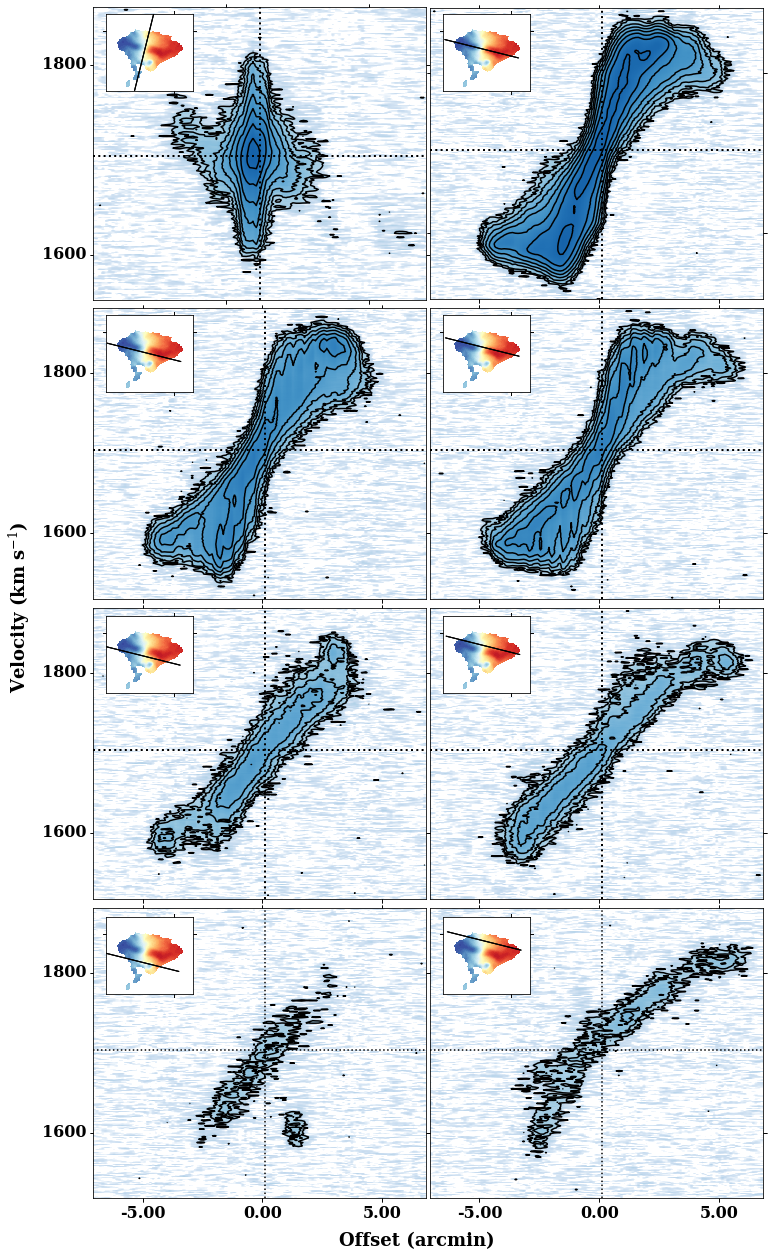}
    \caption{The top left panel shows a bv diagram of UGCA\,250, while the other panels display lv diagrams.
    All panels are at the resolution of 34.3$\arcsec$ $\times$ 25.6$\arcsec$ ({\tt r15t00}, $r =$ 1.5 and $t =$ 0). Contours begin at 3$\sigma$ and increase by factors of 2. Each panel contains an inset indicating the path of
the PV slice on the corresponding moment maps.} 
    \label{fig:lv_maps}
\end{figure*}
\begin{figure}
    \centering
    \includegraphics[width=0.99\linewidth]{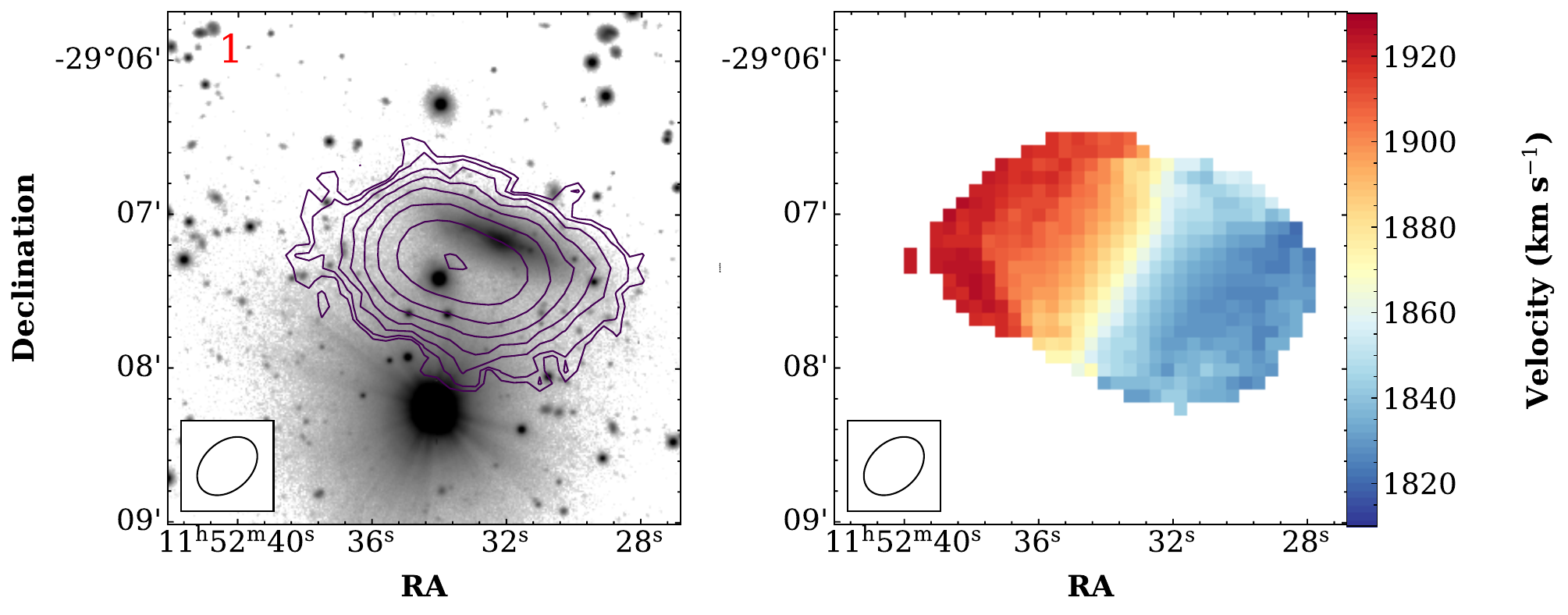}
    \includegraphics[width=0.99\linewidth]{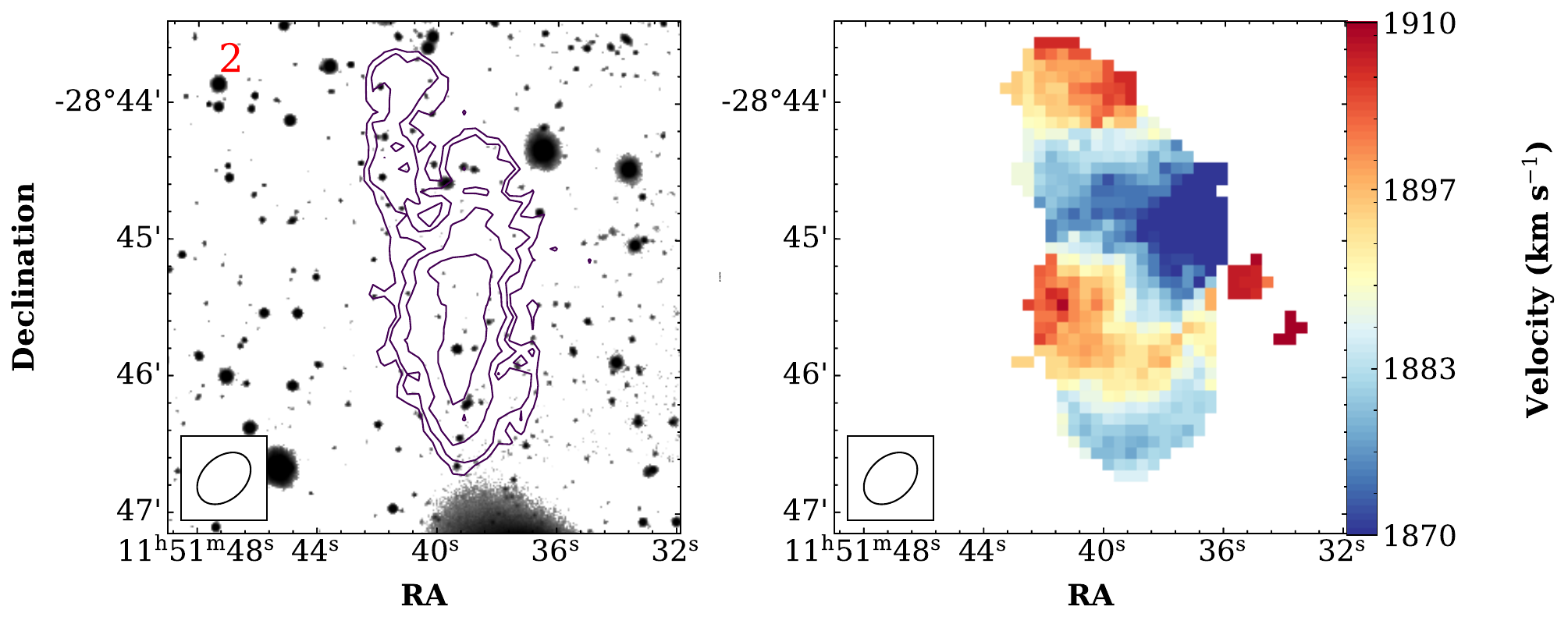}
    \includegraphics[width=0.99\linewidth]{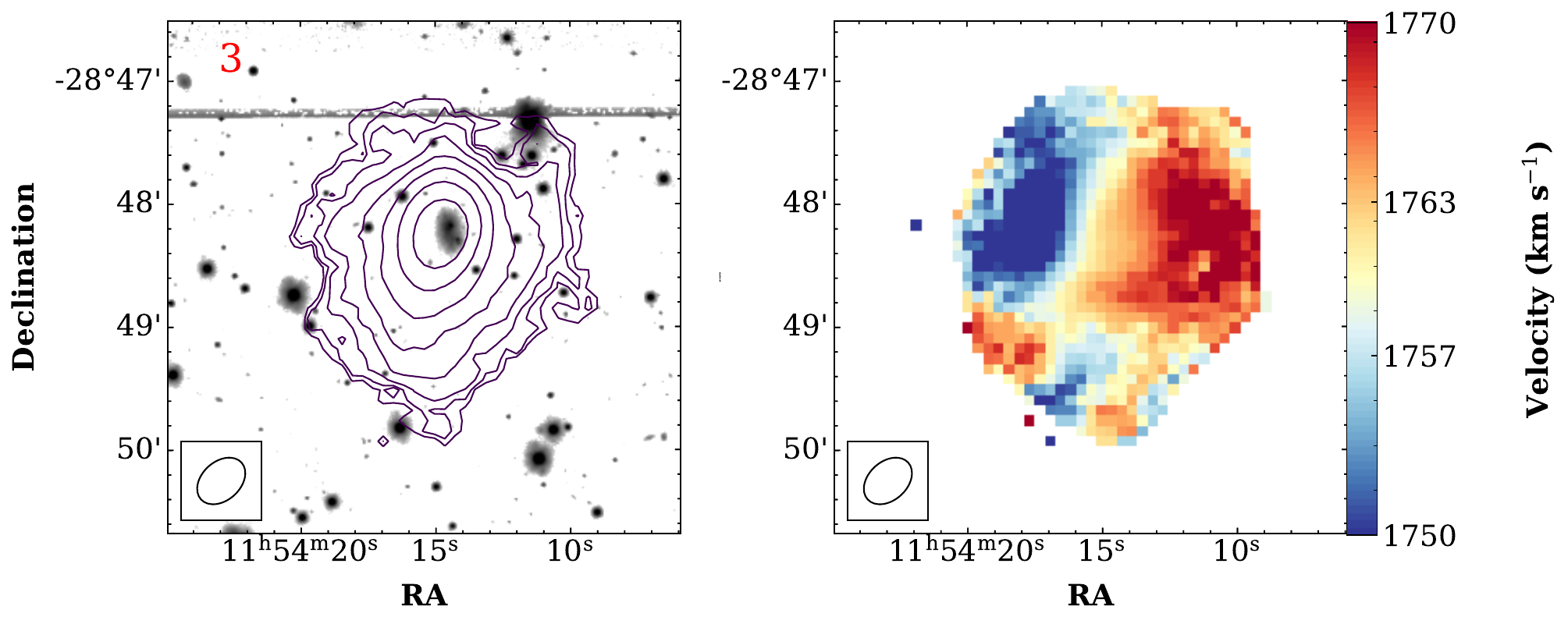}
    \includegraphics[width=0.99\linewidth]{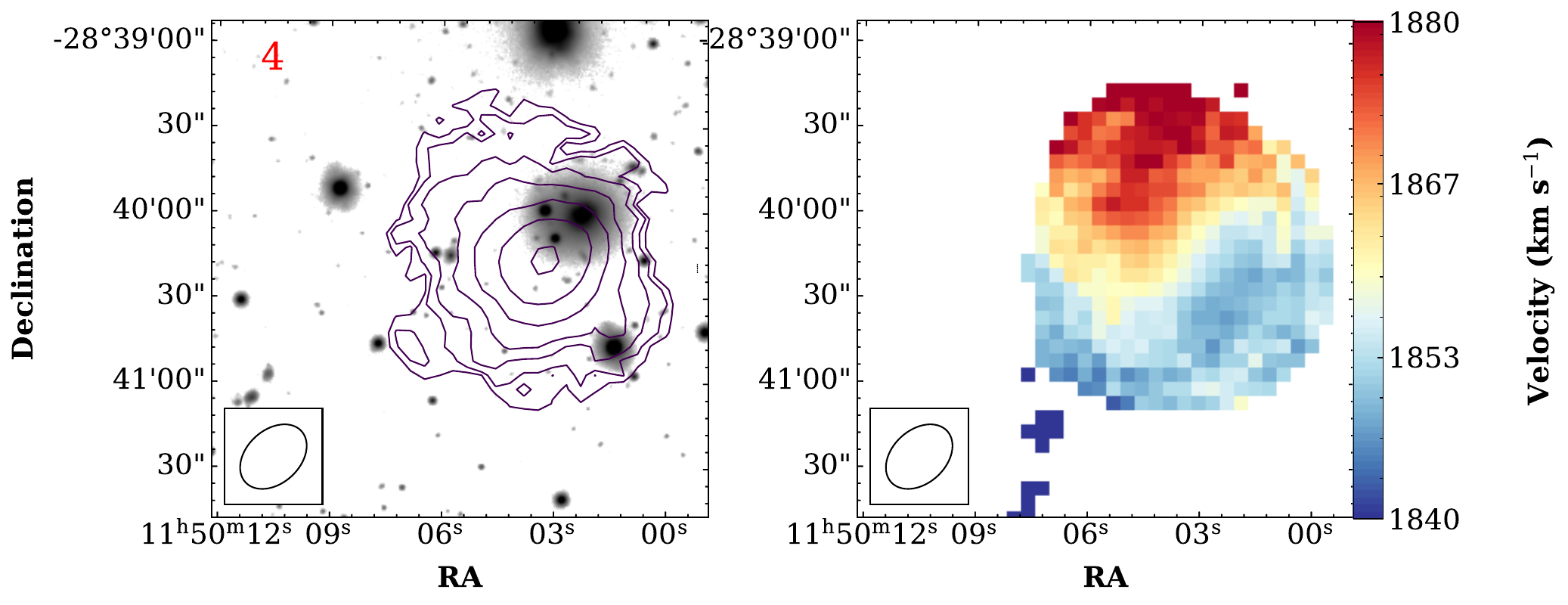}
    \caption{The \HI intensity map (moment 0) and velocity field (moment 1) of satellite galaxies. The H{\sc i} contours are at levels of 1.2$\times$10$^{19}$ cm$^{-2} \times 2^{n}$, n = 0,1,2,3 ..}
    \label{fig:sat1}
\end{figure}
\begin{figure}
    \centering
    \includegraphics[width=0.99\linewidth]{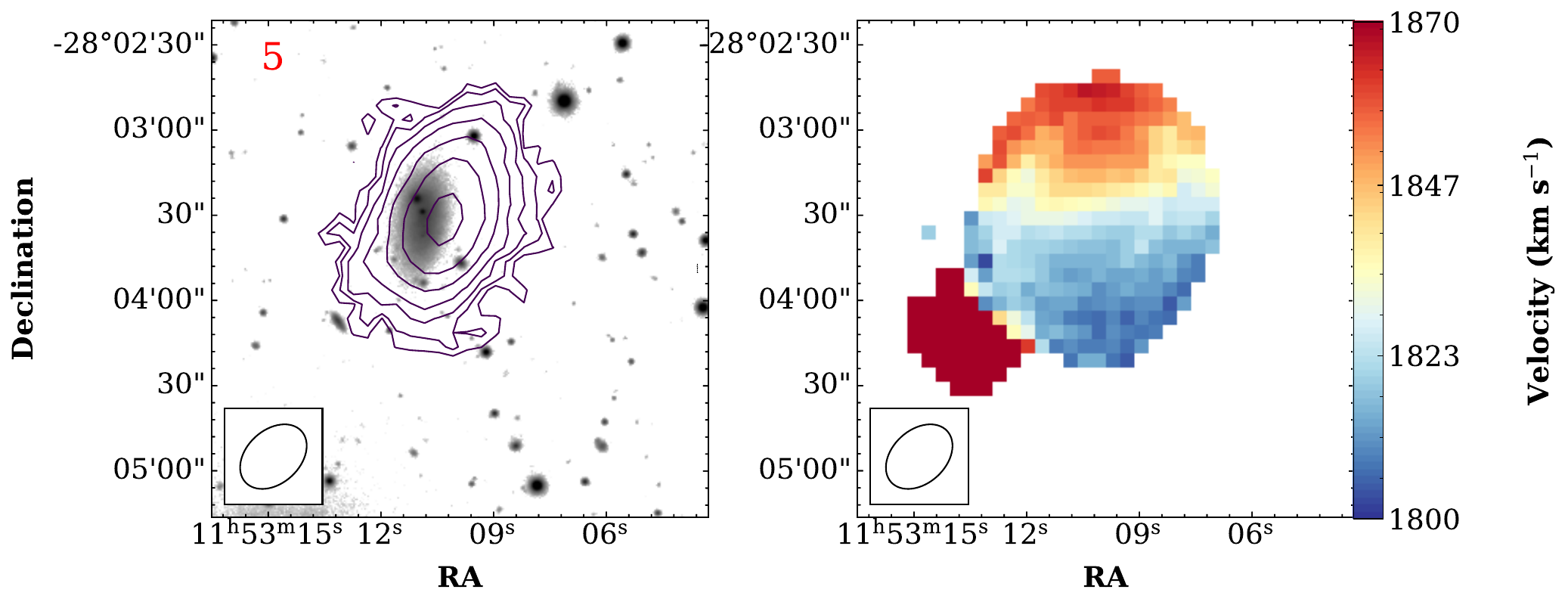}
    \includegraphics[width=0.99\linewidth]{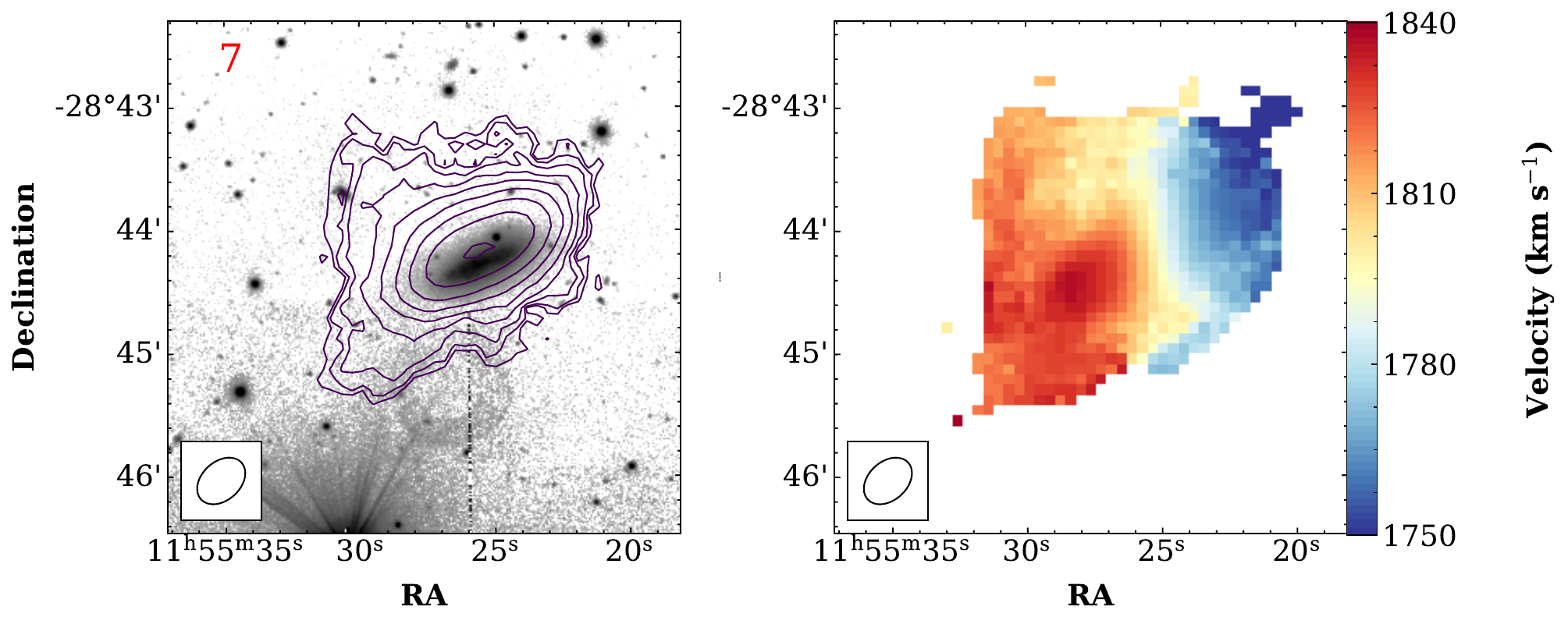}
    \includegraphics[width=0.99\linewidth]{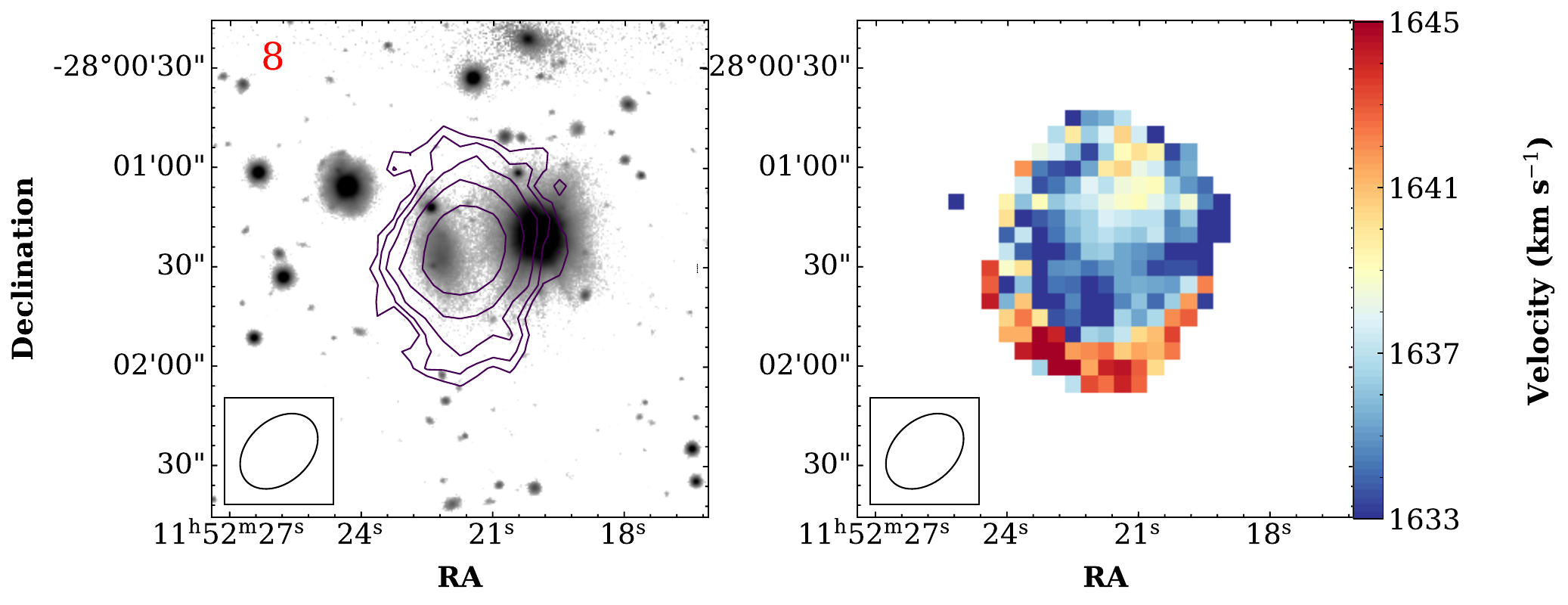}
    \includegraphics[width=0.99\linewidth]{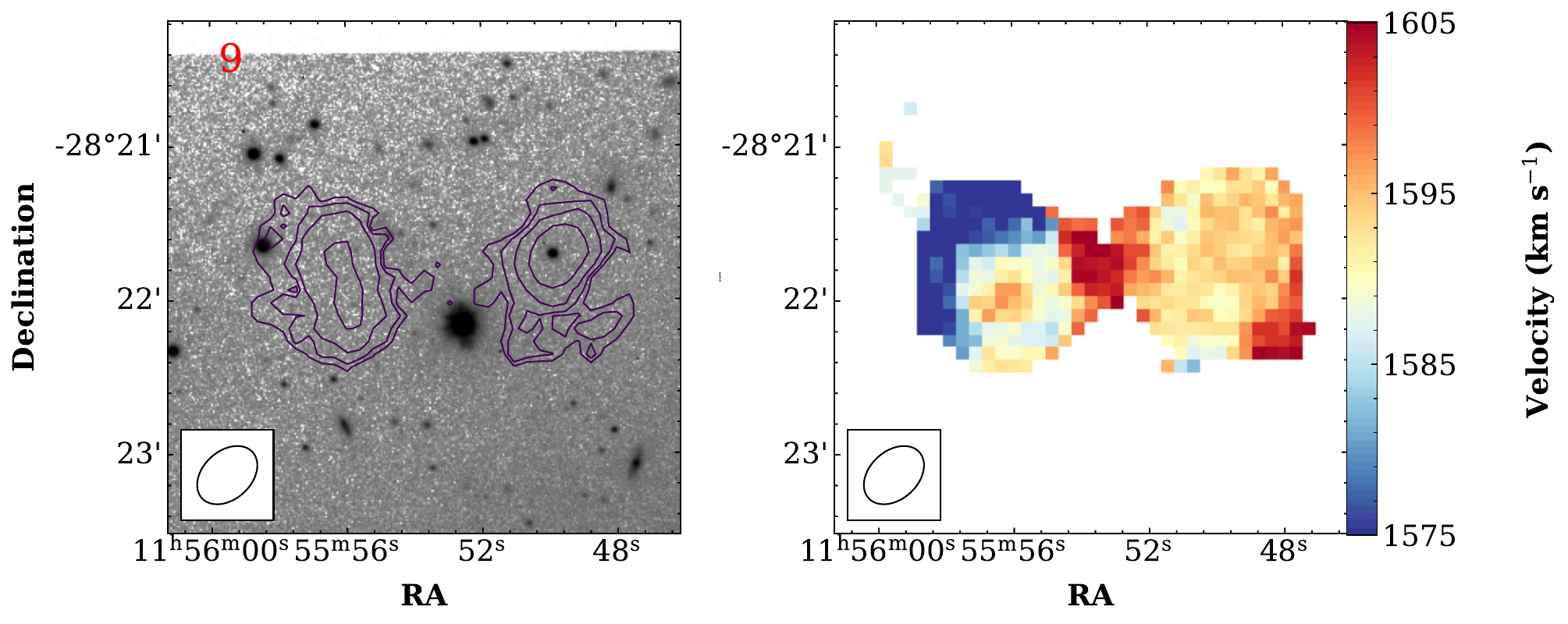}
    \includegraphics[width=0.99\linewidth]{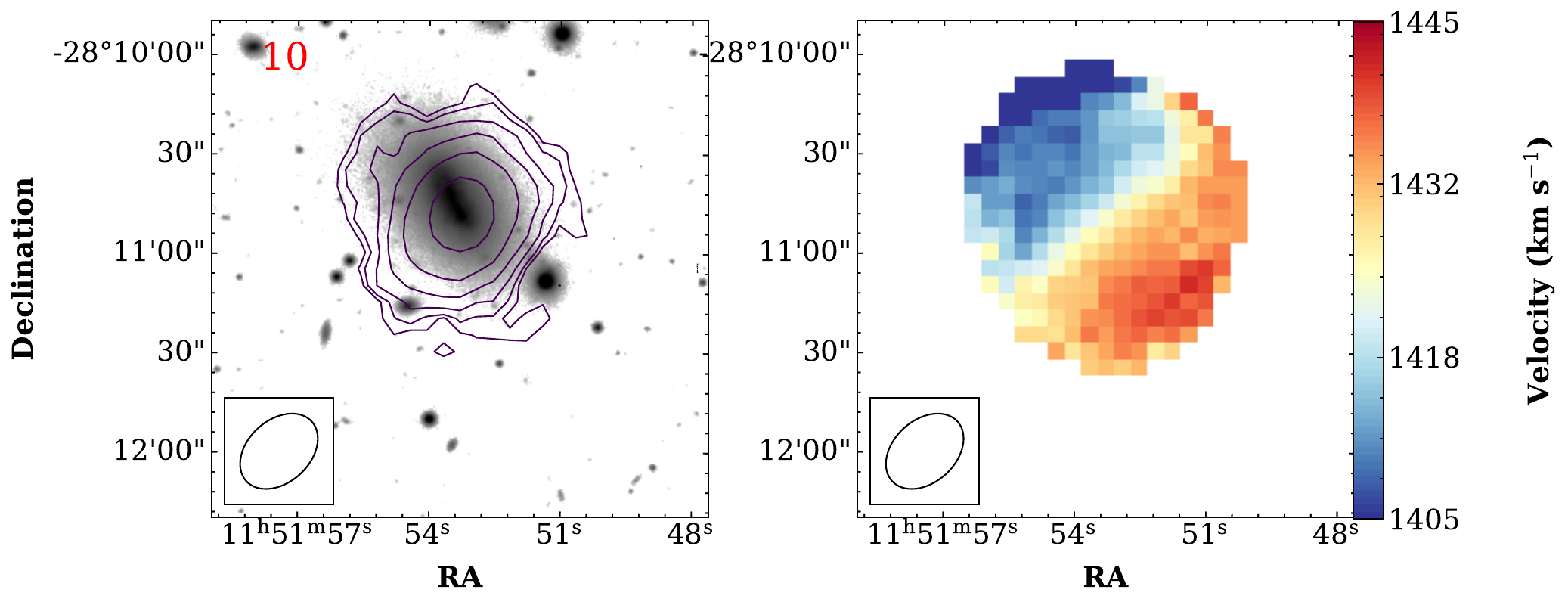}
    \caption{The \HI intensity map (moment 0) and velocity field (moment 1) of satellite galaxies. The H{\sc i} contours are at levels of 1.2$\times$10$^{19}$ cm$^{-2} \times 2^{n}$, n = 0,1,2,3 ..}
    \label{fig:sat2}
\end{figure}

In this appendix,  we present the 
bv and lv diagrams of UGCA 250 (see Fig. \ref{fig:PV_min_1_2} and Fig. \ref{fig:lv_maps}). The bv diagrams, extracted along the minor axis with slices at 30$\arcsec$ intervals on both the approaching and receding halves, clearly display features such as the cloud and the filament extending from the disk. For the lv diagrams along the major axis, an inset is included that shows the slice across the moment 1 map used to generate the lv diagram.  In the lv diagrams along the major axis, the galaxy appears more extended toward the northern, receding half, further underscoring an asymmetry in its structure. Further, we show the moment 0 and moment 1 maps of individual satellite galaxies in Figs \ref{fig:sat1} and \ref{fig:sat2}. 


\bsp	
\label{lastpage}
\end{document}